\documentclass[preprint2]{aastex63}

\shorttitle{Core Growth in IRDC}
\shortauthors{Kong et al.}

\begin{document}

\title{Evidence of Core Growth in the Dragon Infrared Dark Cloud: A Path for Massive Star Formation}

\author[0000-0002-8469-2029]{Shuo Kong}
\affil{Steward Observatory, University of Arizona, Tucson, AZ 85719, USA}

\author[0000-0001-5653-7817]{H\'ector G. Arce}
\affil{Department of Astronomy, Yale University, New Haven, CT 06511, USA}

\author{Yancy Shirley}
\affil{Steward Observatory, University of Arizona, Tucson, AZ 85719, USA}

\author{Colton Glasgow}
\affil{College of Science for Physics, Rochester Institute of Technology, Rochester, NY 14623, USA}

\begin{abstract}
A sample of 1.3 mm continuum cores in the Dragon infrared dark
cloud (also known as G28.37+0.07 or G28.34+0.06) is 
analyzed statistically. Based on their association with
molecular outflows, the sample is divided into protostellar
and starless cores. Statistical tests suggest that
the protostellar cores are more massive than the starless
cores, even after temperature and opacity biases
are accounted for. We suggest that the mass difference
indicates core mass growth since their formation. 
The mass growth implies that massive star formation may not
have to start with massive prestellar cores, 
depending on the core mass growth rate. 
Its impact on the relation between core mass
function and stellar initial mass function is to be 
further explored.
\end{abstract}

\keywords{stars: formation}

\section{Introduction}

How massive stars ($> 8$ M$_\odot$) gain their mass has been debated for decades. It is difficult to accumulate a large amount of mass at the scale of a core ($\sim$0.1 pc) in order to form a massive star before the core fragments and forms a few low mass stars instead. For instance, assuming a core-to-star efficiency of 50\%, a massive star of 50 M$_\odot$ would require a core of 100 M$_\odot$, if the core is the only source of the mass of the forming star. However, under typical conditions (e.g., a sound speed $c_s$ = 0.2 km s$^{-1}$ and a gas number density $n_H$ = 10$^5$ cm$^{-3}$), the Jeans mass is only $\sim$0.2 M$_\odot$. It is unclear how a core with more than 100 Jeans masses would survive the fragmentation that can give rise to hundreds of low-mass cores instead. 

Observational researchers have been searching for massive prestellar cores \citep[e.g.,][]{2017ApJ...834..193K,2018sf2a.conf..311L,2019A&A...626A.132M,2019ApJ...886...36S}; however, no convincing candidates
have been found.
Moreover, massive stars usually come in pairs \citep[or multiples,][]{2013ARA&A..51..269D}, meaning the core would need to at least have double the amount of mass. Therefore, it is challenging to picture that only the mass enclosed in a core is responsible for the formation of massive stars.

More generally speaking, the crux of the matter is the way the mass is assembled. 
It is not yet clear whether the full amount of mass needed to form a massive star is contained in a small volume that is virialized without fragmentation (i.e., the core), that is then followed by gravitational collapse, or if the core contains just enough mass to form a ``seed'' and most of the mass needed to build the massive star is funneled through the core from the surrounding cloud. 

Recently, \citet{2018MNRAS.473.4220L} and \citet{2019MNRAS.485.4686W} showed that filaments could channel material to cores (with or without magnetic fields). This can largely ease the burden of having to accumulate an excessive amount of mass during the prestellar core phase because additional mass can be provided later for the forming massive star during its accretion phase. This is similar to the traditional clump-fed picture \citep{2009MNRAS.400.1775S} in the sense that the mass reservoir is at a much larger scale than the parent core. However, instead of a spherical collapse, the mass is transferred to the core and the protostar through filaments, which is also suggested by a recent numerical study by \citet{2020ApJ...900...82P}. 

A few features of this filament-fed picture have been found recently by other studies. For instance, \citet{2019MNRAS.485.4509L} have shown that the longitudinal filament mass flow causes the protostellar outflow to be preferentially orthogonal to the filament, which was clearly detected in an infrared dark cloud \citep[IRDC,][]{2019ApJ...874..104K}. In addition, recent observations of polarized dust emission in filaments show that magnetic fields are mostly perpendicular to star-forming filaments \citep{2019ApJ...876...42W,2019ApJ...883...95S}, consistent with the picture of the filament-fed accretion shown in \citet{2019MNRAS.485.4509L}\footnote{Note, however, the perpendicular configuration may be disturbed due to protostellar evolution, as seen in \citet{2020ApJ...890...44B}.}. Here, the perpendicular magnetic fields channel material to the filament \citep{2012ApJ...745L..30W}. 

If indeed cores gain additional mass from the hosting filament, their mass should increase over time. Depending on the filament-core accretion rate, this core mass increase may be notable in observations. For instance, infall signatures have been detected toward massive clumps \citep[e.g.,][]{2007ApJ...669L..37W,2010MNRAS.402...73B,2011ApJ...740...40R,2015MNRAS.450.1926H,2018ApJ...861...14C,2018ApJ...862...63C,2018ApJS..235...31Y,2019ApJ...870....5J}. 

In this paper, we investigate the statistical difference between the population of cores with and without protostellar outflows. By comparing the mass and size distribution of the cores in these two populations, we search for evidence of core growth. In the following, we introduce the sample collection in \S\ref{sec:obsalma}. We report our findings in \S\ref{sec:results}. Discussions and conclusions are in \S\ref{sec:disc} and \S\ref{sec:conc}.

\section{Sample Selection}\label{sec:obsalma}

The statistical sample consists of two parts. The first is a sample of cores defined by \citet[][hereafter K19a]{2019ApJ...873...31K} and the second an outflow sample defined by \citet[][hereafter K19b]{2019ApJ...874..104K}. There are two different core samples based on the core-finding algorithm. In addition, the core masses are obtained with two different methods, differing on how the dust temperatures were estimated.

K19a studied the core mass function (CMF) in the Dragon IRDC based on an ALMA 1.3 mm continuum mosaic of the cloud. The mosaic consisted of eighty-six ALMA pointings that covered the majority of the dark cloud. K19a used two methods to dissect the continuum emission. The first core-finding algorithm, which found 280 cores, was developed by K19a and was based on Graph algorithms ({\it astrograph}). The second method used the Dendrogram technique ({\it astrodendro}, \citealt{2008ApJ...679.1338R}) and found 197 continuum cores. The main difference between the two methods was that {\it astrograph} finds more low-mass cores in crowded regions than {\it astrodendro}. See K19a for a more detailed comparison. In this paper, we use the two core samples for independent tests.

K19a derived core masses based on two temperature estimations. First, K19a assumed a constant core dust temperature of 20 K, which is typical in IRDCs \citep{2006A&A...450..569P,2011ApJ...735...64W}. Second, K19a used the NH$_3$-based kinetic temperature map, obtained from observations using the Karl G. Jansky Very Large Array \citep[][hereafter W18]{2018RNAAS...2...52W} to determine the core dust temperature. In this paper, we use the two independent core mass estimations for our analysis.
Following \citet{2018ApJ...853..160C}, K19a adopted a 1.3 mm opacity ($\kappa_d=0.899~\rm cm^2g^{-1}$, fiducial value) based on the moderately coagulated thin ice mantle model from \citet[][hereafter OH94]{1994A&A...291..943O}. Because we are comparing masses of starless cores and protostellar cores, we will also address the possible
opacity difference between the two core populations, as the latter may have systematically higher opacities due to coagulation. 

In total, K19b identified 62 {\it astrograph} cores with CO and/or SiO outflows. However, 6 of them were not in the core catalog because K19a searched cores within a primary-beam response of 0.5 for the CMF study. K19b searched protostellar outflows originating in cores within a primary-beam response of 0.2, giving rise to 6 more detections. Here we use the outflow sample excluding the 6 additional cores in order to be consistent with the K19a core sample. Of the sample of 56 {\it astrograph} protostellar cores, 51 of them have {\it astrodendro} counterparts. Five of the 56 {\it astrograph} cores were not found by {\it astrodendro}. Hence, in total, 51 of the 197 {\it astrodendro} cores are defined as protostellar.


\section{Molecular Line Data and Gas Kinematics}\label{sec:line}

To estimate the core virial status, 
we utilize the molecular line data 
from ALMA projects 2013.1.00183.S
and 2015.1.00183.S. In particular,
we use the C$^{18}$O(3-2), DCO$^+$(3-2), 
N$_2$D$^+$(3-2), and DCN(3-2) line cubes to
derive the kinematic information for the cores.
We jointly cleaned the two data sets with 
natural weighting to reach a final synthesized 
beam of $\sim$0.8\arcsec. We adopted a cube channel
width of 0.2 km s$^{-1}$ to balance between
the spectral resolution and the sensitivity 
per channel. The final sensitivity is $\sim$
8 mJy beam$^{-1}$ per 0.2 km s$^{-1}$ channel. 
In \S\ref{app:kin}, we describe in detail how
we fit the core spectra.

\section{Results and Analyses}\label{sec:results}

\subsection{Mass Difference between Core Populations}

\begin{figure*}[htb!]
\centering
\epsscale{1.1}
\plotone{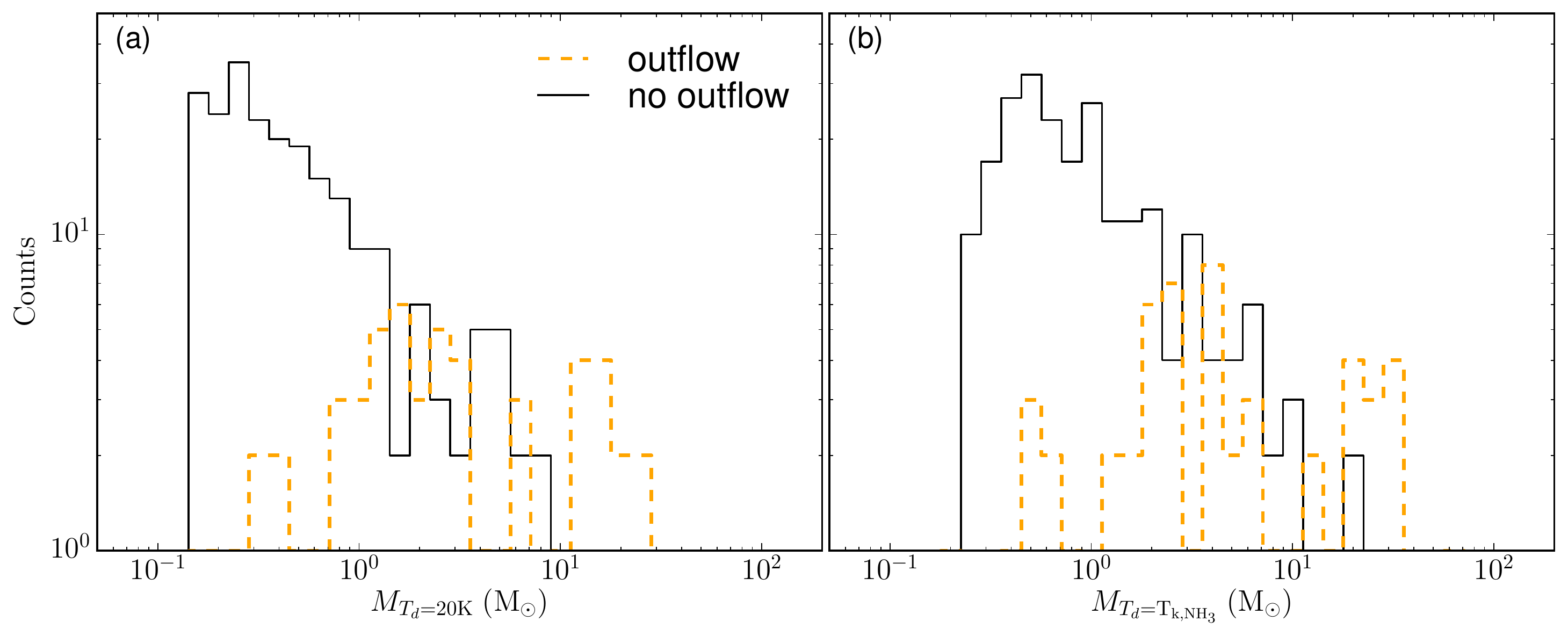}\\
\plotone{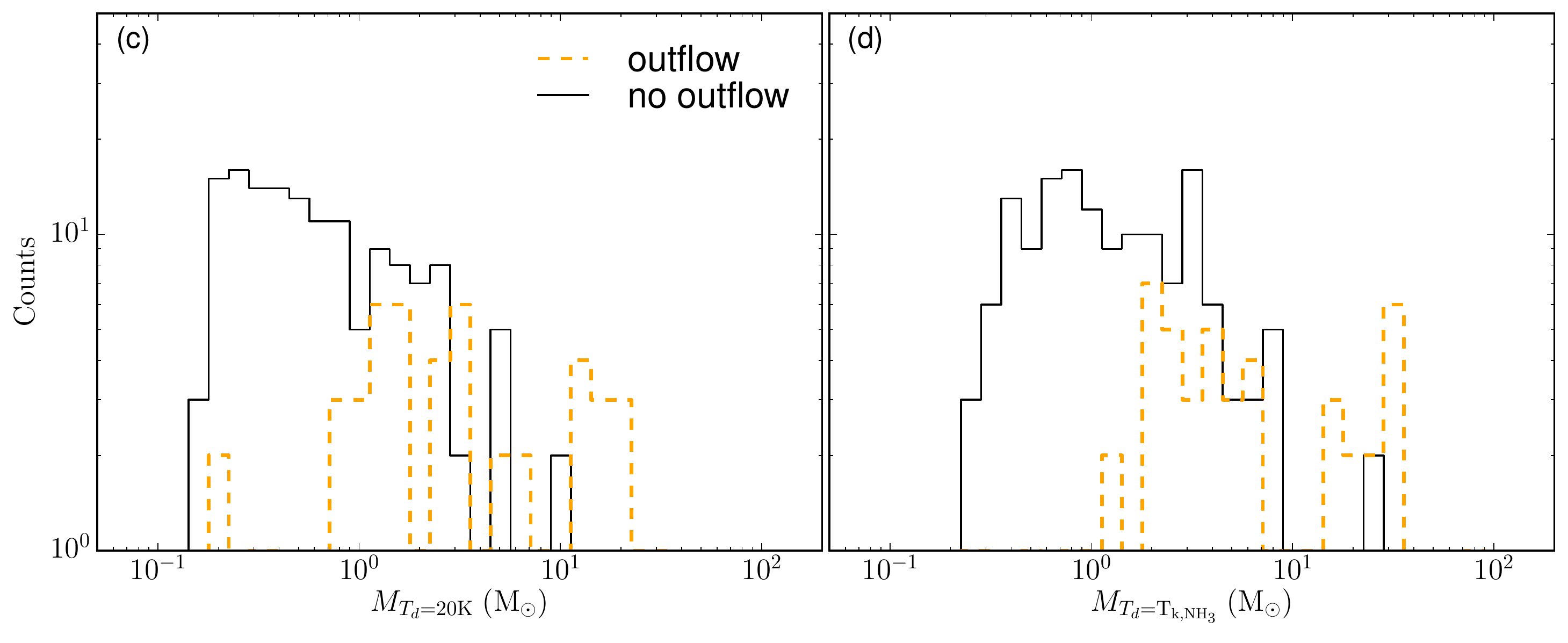}
\caption{
Core mass histograms for {\it astrograph} and {\it astrodendro} samples from K19a. 
{\bf (a):} Core mass computed based on a constant 
dust temperature of 20 K. The orange dashed histogram
shows the protostellar cores with outflows. The black
histogram shows the cores without outflows. 
{\bf (b):} Same as panel (a), but the core mass is
computed based on the NH$_3$ kinetic temperature.
{\bf (c):} Same as panel (a), but for the 
{\it astrodendro} core sample.
{\bf (d):} Same as panel (b), but for the 
{\it astrodendro} core sample.
\label{fig:mass}}
\end{figure*}

Figures \ref{fig:mass}(a)(b) show the mass histograms for the {\it astrograph} core sample. Core masses in panel (a) are computed assuming the constant dust temperature of 20 K. The median value for protostellar core masses is 2.1 M$_\odot$; for starless cores is 0.37 M$_\odot$. Panel (a) shows that protostellar cores tend to be more massive than starless cores. Using the NH$_3$ kinetic temperatures for each core does not change the result, as shown in panel (b). Here, the median value for protostellar core masses is 3.7 M$_\odot$; for starless cores is 0.74 M$_\odot$.

To statistically confirm the result, we carry out a Mann-Whitney U test \citep{mann1947} for the samples. The test is a nonparametric test for the ranking between two samples. Here we test whether the cores with outflows are more massive than those without outflows. For panel (a), the null hypothesis that the cores with outflows are equally or less massive than the cores without outflows can be rejected with a confidence greater than 99.99\% (p-value $\ll$ 0.01\%). The same conclusion applies to the core sample in panel (b). These results suggest that the protostellar cores tend to be more massive than starless cores in the Dragon IRDC.

Figures \ref{fig:mass}(c)(d) show the same analysis for the {\it astrodendro} cores. Again, regardless of the temperature estimation, the cores with outflows tend to be more massive than those without outflows. Here, the Mann-Whitney U tests yield the same conclusion as before, that the null hypothesis that the cores with outflows are equally or less massive than the cores without outflows can be rejected with a confidence greater than 99.99\% for both panels, suggesting that {\it astrodendro} cores with outflows (protostellar) are, in general, more massive than those without outflows (starless).

\begin{figure*}[htb!]
\centering
\epsscale{1.1}
\plotone{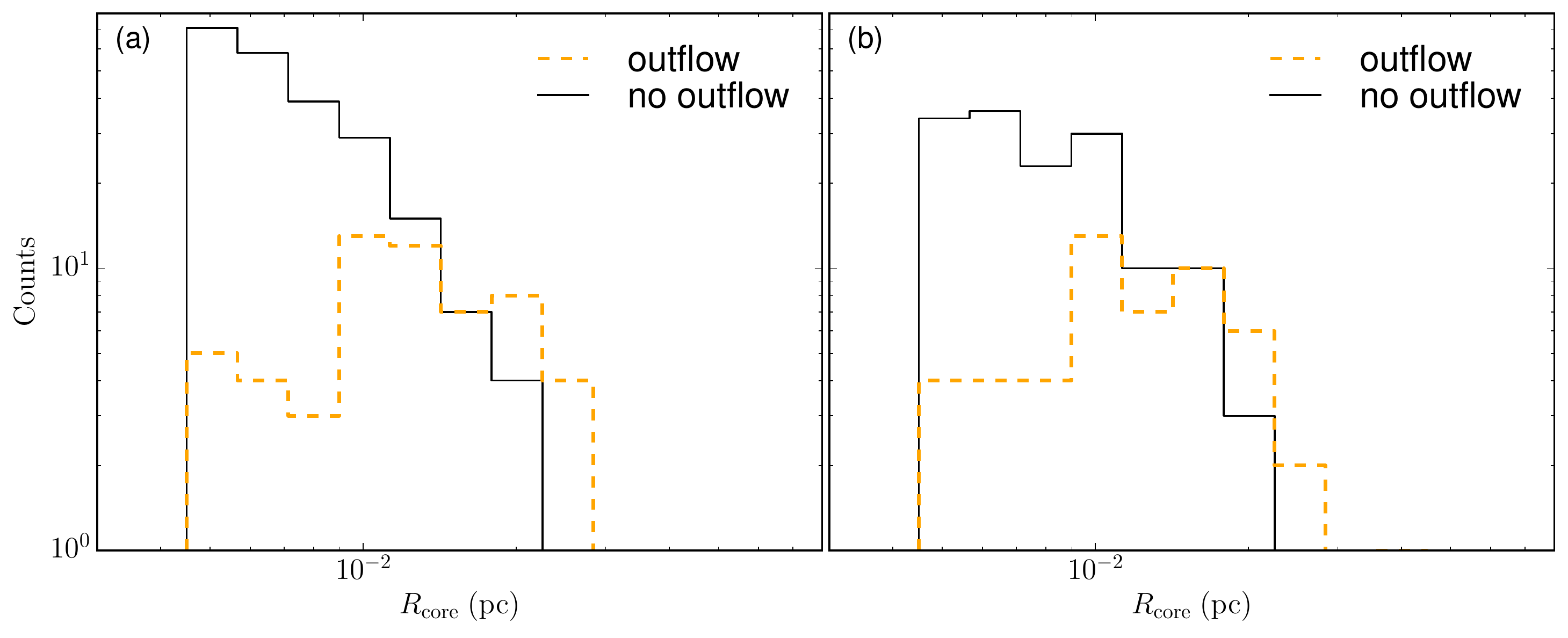}
\caption{
Core radius histograms. {\bf (a):} {\it astrograph} core
radii in parsec. The orange dashed histogram is for the
protostellar cores. The black histogram is for the 
starless cores. {\bf (b):} Same as (a), but for 
{\it astrodendro} cores.
\label{fig:size}}
\end{figure*}

The reason that the protostellar cores are more massive than the starless cores is because, overall, protostellar cores are more likely to have larger sizes than starless cores, while the two populations show no difference in their average density distribution. Here we assume that the core  radius is the equivalent radius of a circle that has the same area of the core area $R_{\rm core}\equiv(A_{\rm core}/\pi)^{0.5}$, where the core area $A_{\rm core}$ is from the core definition in K19a.

In Figure \ref{fig:size}, we show histograms of the core radii. Panel (a) shows the results for {\it astrograph} cores while panel (b) is for {\it astrodendro} cores (K19a). The histograms show that radii of the protostellar cores tend to be larger than the starless cores. For both {\it astrograph} and {\it astrodendro} samples, the null hypothesis that the cores with outflows are equal or smaller than the cores without outflows can be rejected with a confidence greater than 99.99\%.

\begin{figure*}[htb!]
\centering
\epsscale{1.1}
\plotone{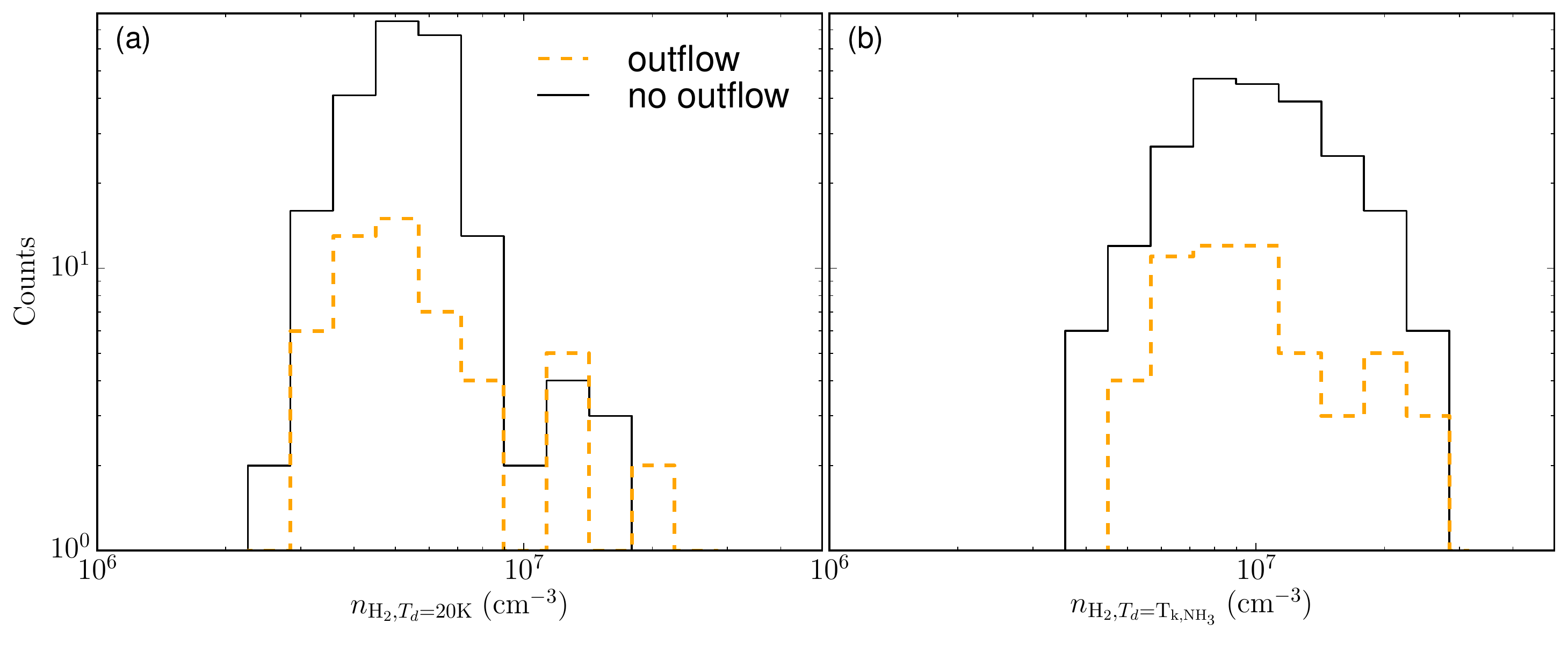}\\
\plotone{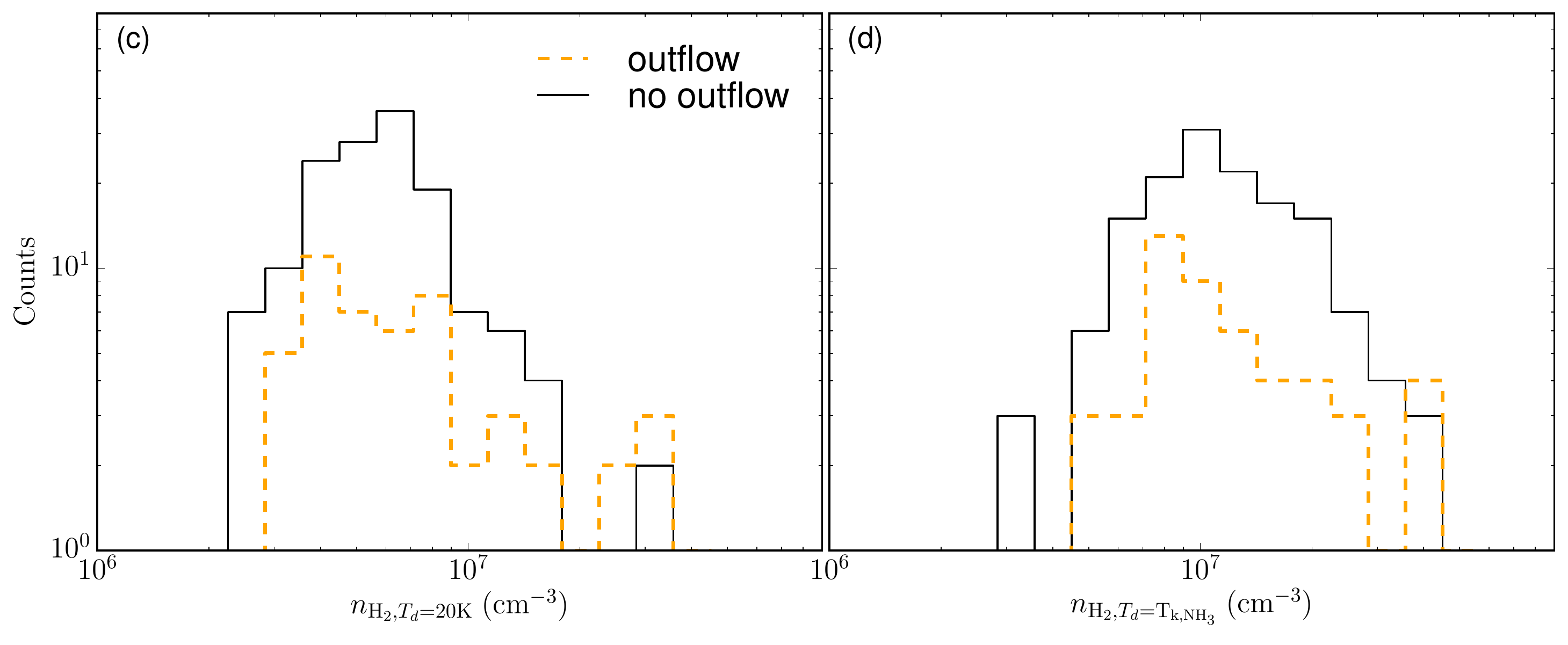}
\caption{
Average core density histograms for the {\it astrograph} sample. 
{\bf (a):} Core density based on core mass from Figure
\ref{fig:mass}(a). {\bf (b):} Core density based on
core mass from Figure \ref{fig:mass}(b).
{\bf (c):} Core density based on core mass from Figure
\ref{fig:mass}(c). {\bf (d):} Core density based on
core mass from Figure \ref{fig:mass}(d).
\label{fig:dens}}
\end{figure*}

Figures \ref{fig:dens}(a)(b) show the distribution of average core densities for {\it astrograph} cores with different temperature assumptions (as indicated by the x-axis labels). The average density is simply computed as the core mass divided by the core volume. The core radius is from Figure \ref{fig:size}. Panels (c) and (d) are for {\it astrodendro} cores.

The protostellar cores do not show systemically higher average density than the starless cores. Again, we apply the Mann-Whitney U test (but now two-sided). For {\it astrograph} cores, the null hypothesis that the protostellar  core average density is either less than or greater than the starless core average density is not rejected with a high confidence (p-value 0.90 for the constant temperature case and 0.49 for the NH$_3$ gas temperature case). The same conclusion applies to the {\it astrodendro} cores (p-value 0.22 for the constant temperature case and 0.68 for the NH$_3$ gas temperature case). These suggest that the average core density is not very different between the protostellar cores and the starless cores. 

At this point, a plausible physical picture is that protostellar cores, emerging earlier in the IRDC, have grown in size and mass over their relatively longer lifetime, while keeping their density roughly invariant. The core density was probably inherited from local environments that were determined by other physical processes over the entire cloud.

\begin{figure*}[htb!]
\centering
\epsscale{0.55}
\plotone{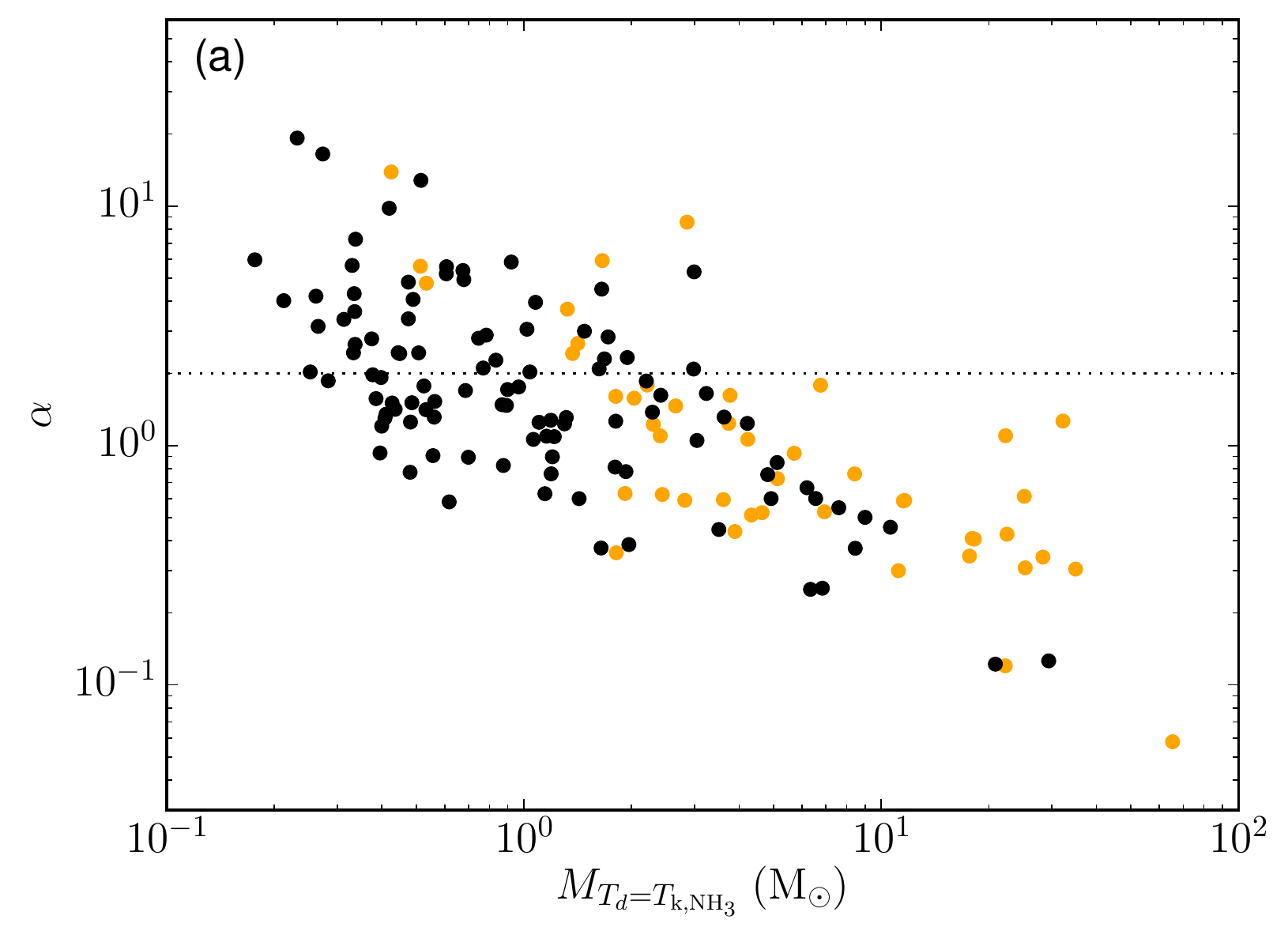}
\plotone{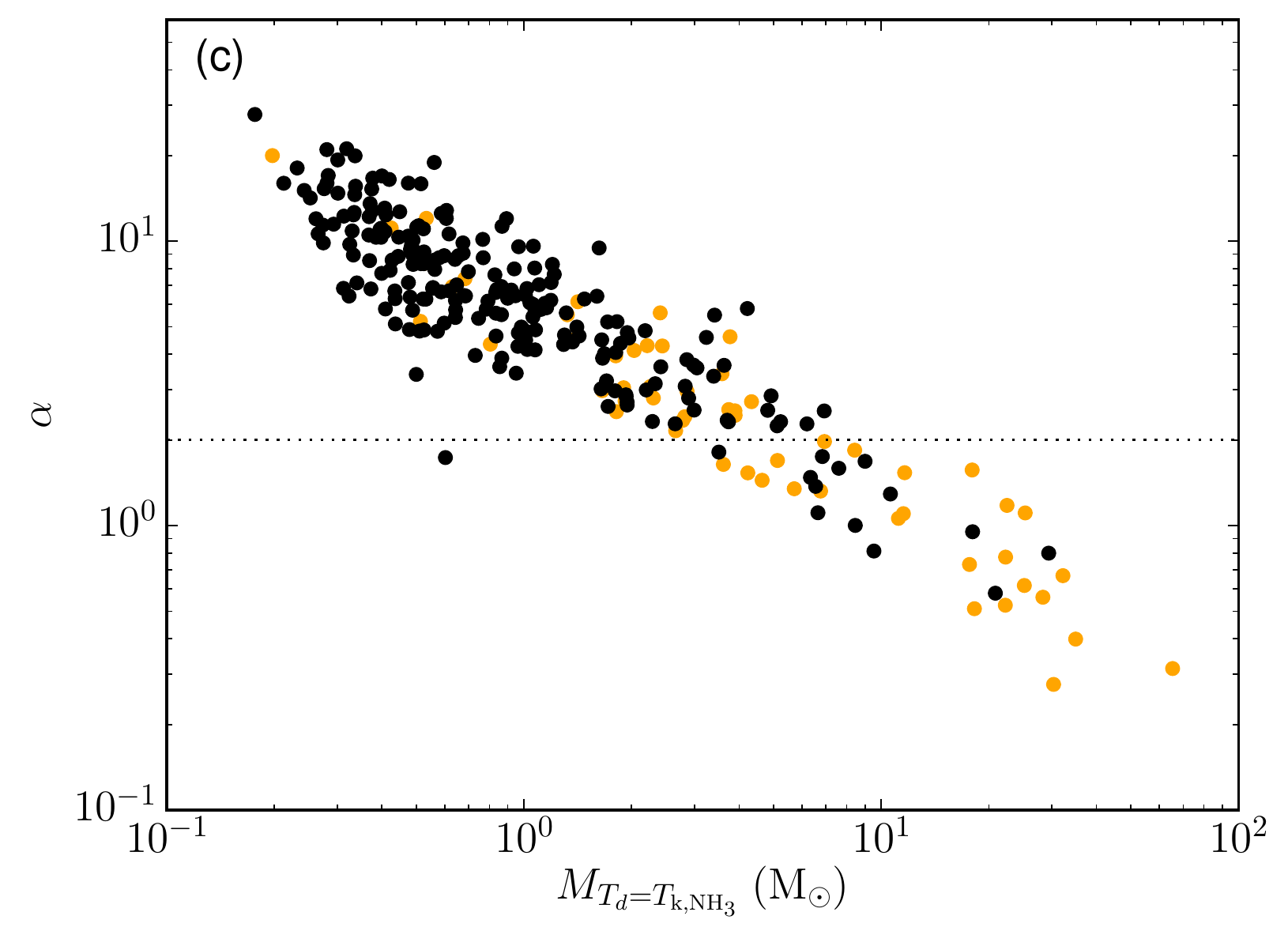}\\
\plotone{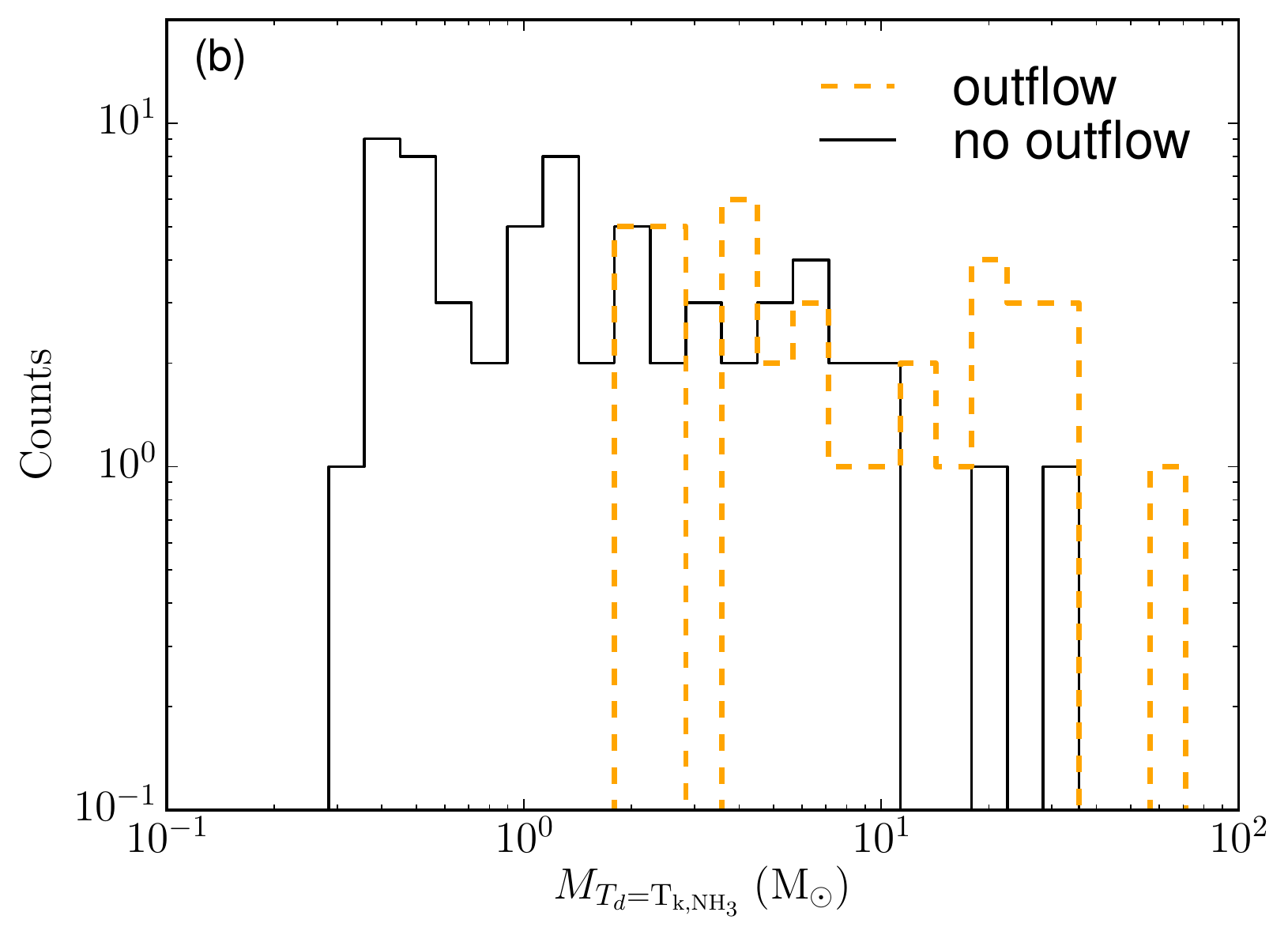}
\plotone{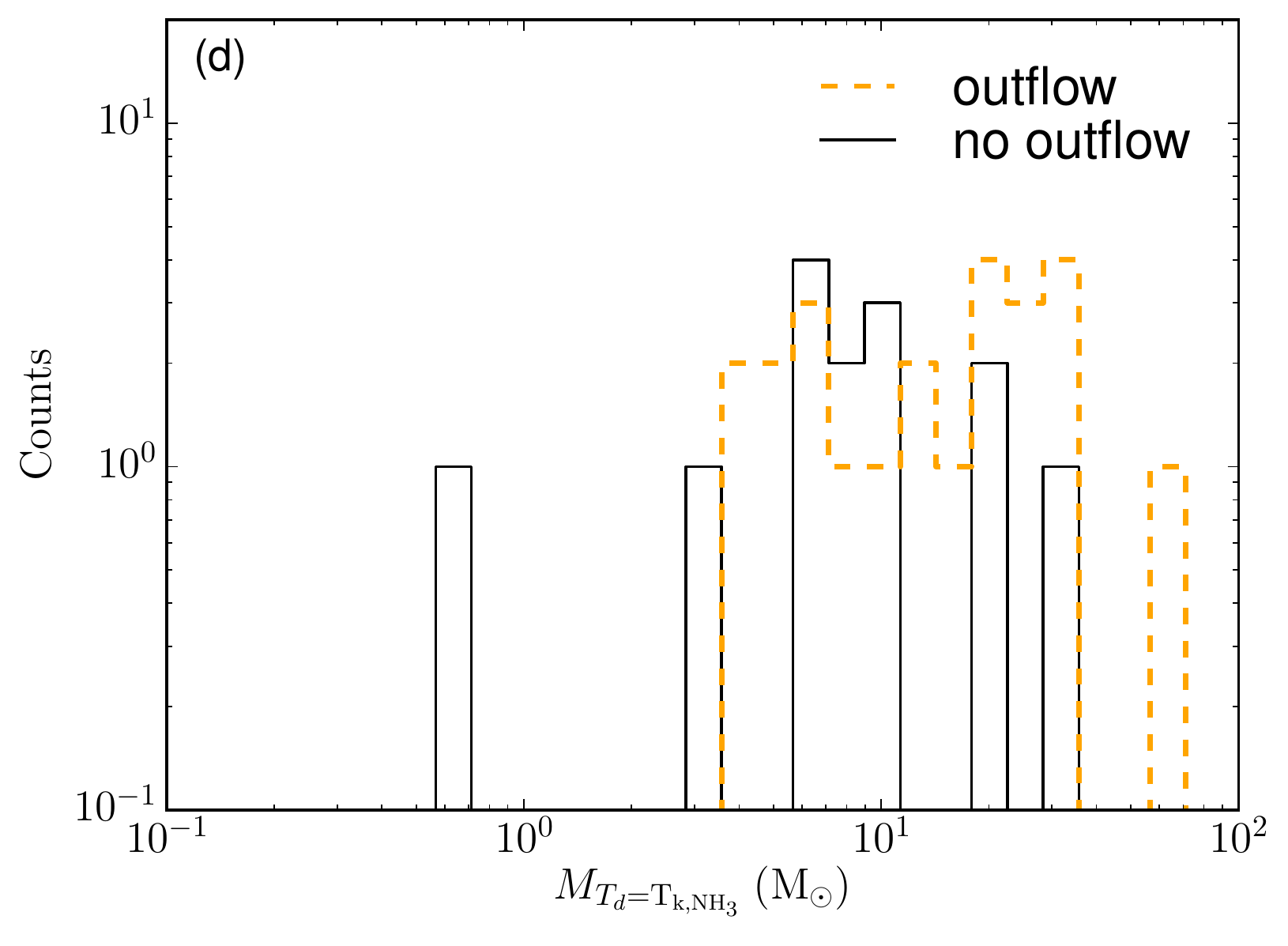}
\caption{
{\bf (a):} Virial parameter versus core mass
for the {\it astrograph} cores. The core velocity
dispersion is from  column (10) of 
Table \ref{tab:kin}. The core mass is based on
the NH$_3$ gas kinetic temperature. The horizontal
dotted line is at $\alpha=2$.
{\bf (b):} Core mass histograms for protostellar
and starless cores with $\alpha<2$ based on panel (a).
{\bf (c):} Same as panel (a), but the
core velocity dispersion is from the column (11)
of Table \ref{tab:kin}. The core mass is based on
the NH$_3$ gas kinetic temperature.
{\bf (d):} Core mass histograms for protostellar
and starless cores with $\alpha<2$ based on panel (c).
\label{fig:alpha}}
\end{figure*}

In the appendix, we derive the core velocity dispersion with the ALMA molecular line data (mainly from C$^{18}$O, see \S\ref{app:kin}) as well as the public NH$_3$ data from W18. This time, we focus on the {\it astrograph} cores.
With the dispersion, we compute the virial parameter ($\alpha=5\sigma_{\rm tot}^2R/GM$) for each core. Figures \ref{fig:alpha}(a)(c) show the results. 

In Figure \ref{fig:alpha}(a), some of the protostellar cores have $\alpha>2$, which indicates that the cores are not gravitationally bound. Massive cores tend to have lower $\alpha$ values. The same behavior is seen in Figure \ref{fig:alpha}(c), but in this plot more cores have $\alpha>2$. This difference is due to the typically larger values of the NH$_3$ velocity dispersion compared to the dispersion measured using the ALMA line data (compare columns (10) and (11) of Table \ref{tab:kin}). As we mentioned earlier, the ALMA data 
filters large-scale structures
while the W18 beam size is larger than the cores. The fact that these two datasets probe different scales 
very likely gives rise to the difference in velocity dispersion observed in these two lines, thus the contrast in the virial parameter derived from these.

In Figure \ref{fig:alpha}(b), we show the distribution of core mass for cores with $\alpha<2$ (i.e., which are gravitationally bound) for both starless and protostellar cores. 
From these histograms, we can see that, in general, protostellar cores are more massive than starless cores. Again, this is strongly supported by the Mann-Whitney U test with a confidence $>$ 99.99\%. 
In Figure \ref{fig:alpha}(d), where we show 
gravitationally bound cores with velocity dispersion based on the NH$_3$ observations,
the total number of  cores is smaller than in Figure \ref{fig:alpha}(b). 
Yet, the protostellar cores are still likely to be more massive than the starless cores with a confidence of 97\%.

\subsection{Potential Biases from Temperature}

\begin{figure}[htb!]
\centering
\epsscale{1.}
\plotone{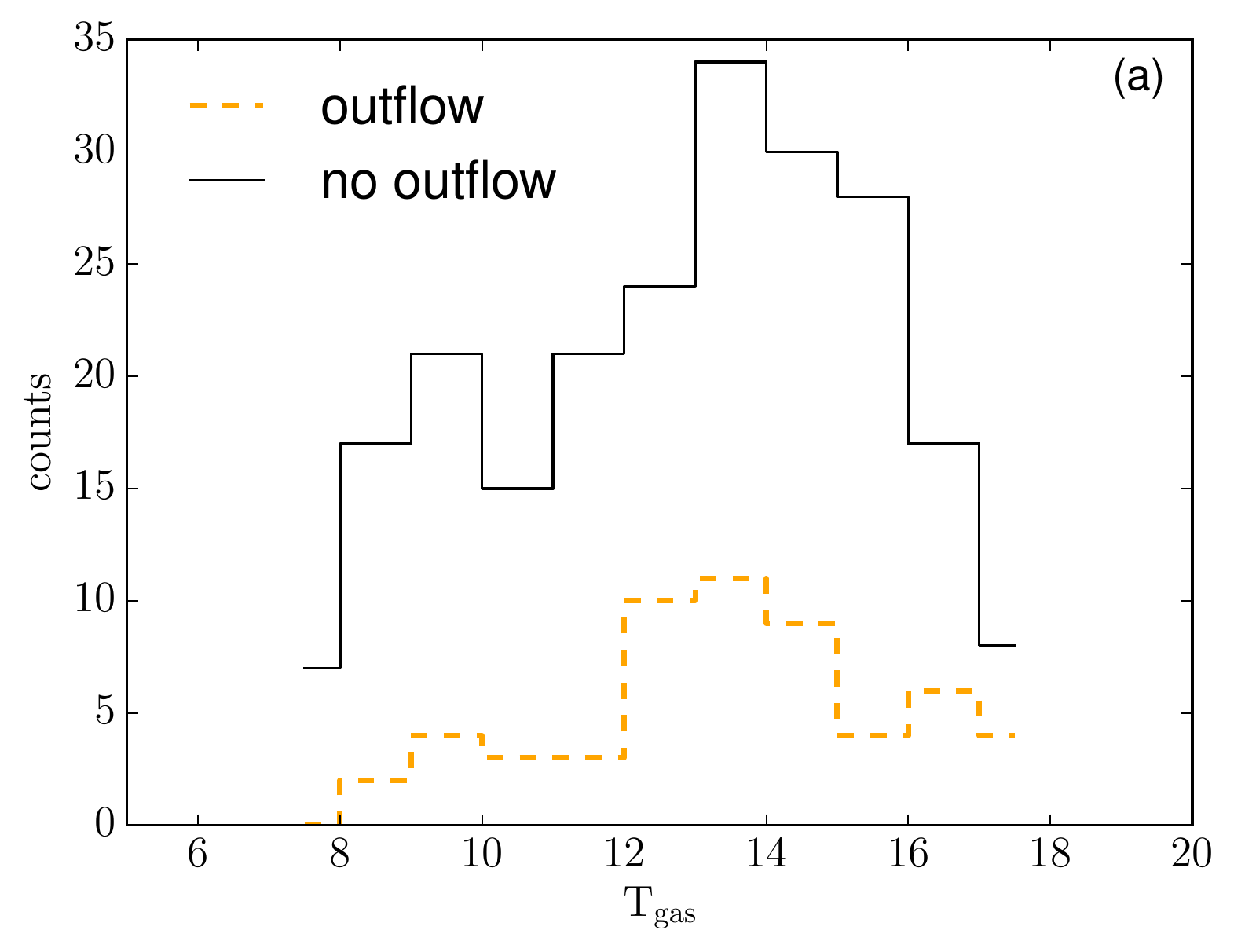}\\
\plotone{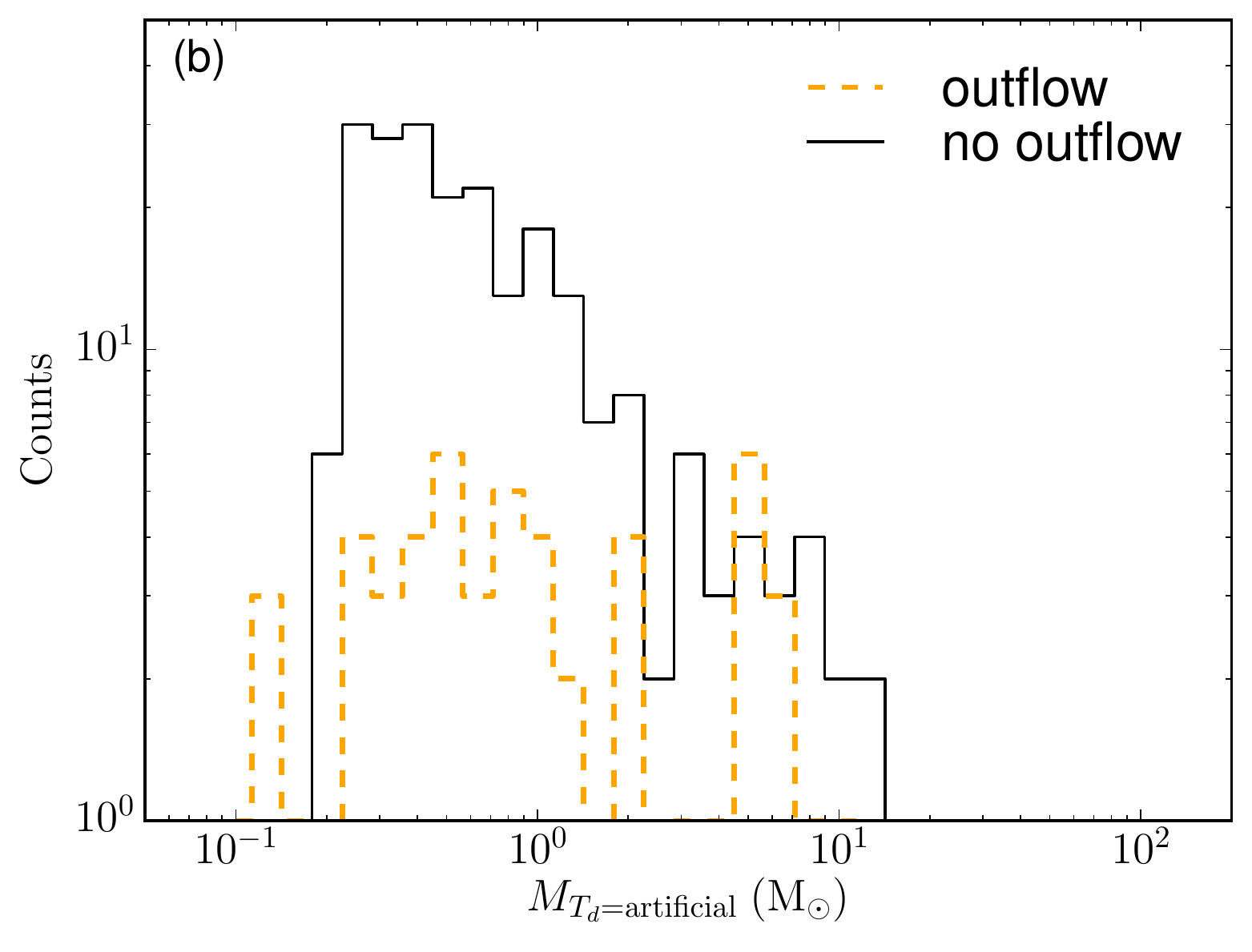}\\
\plotone{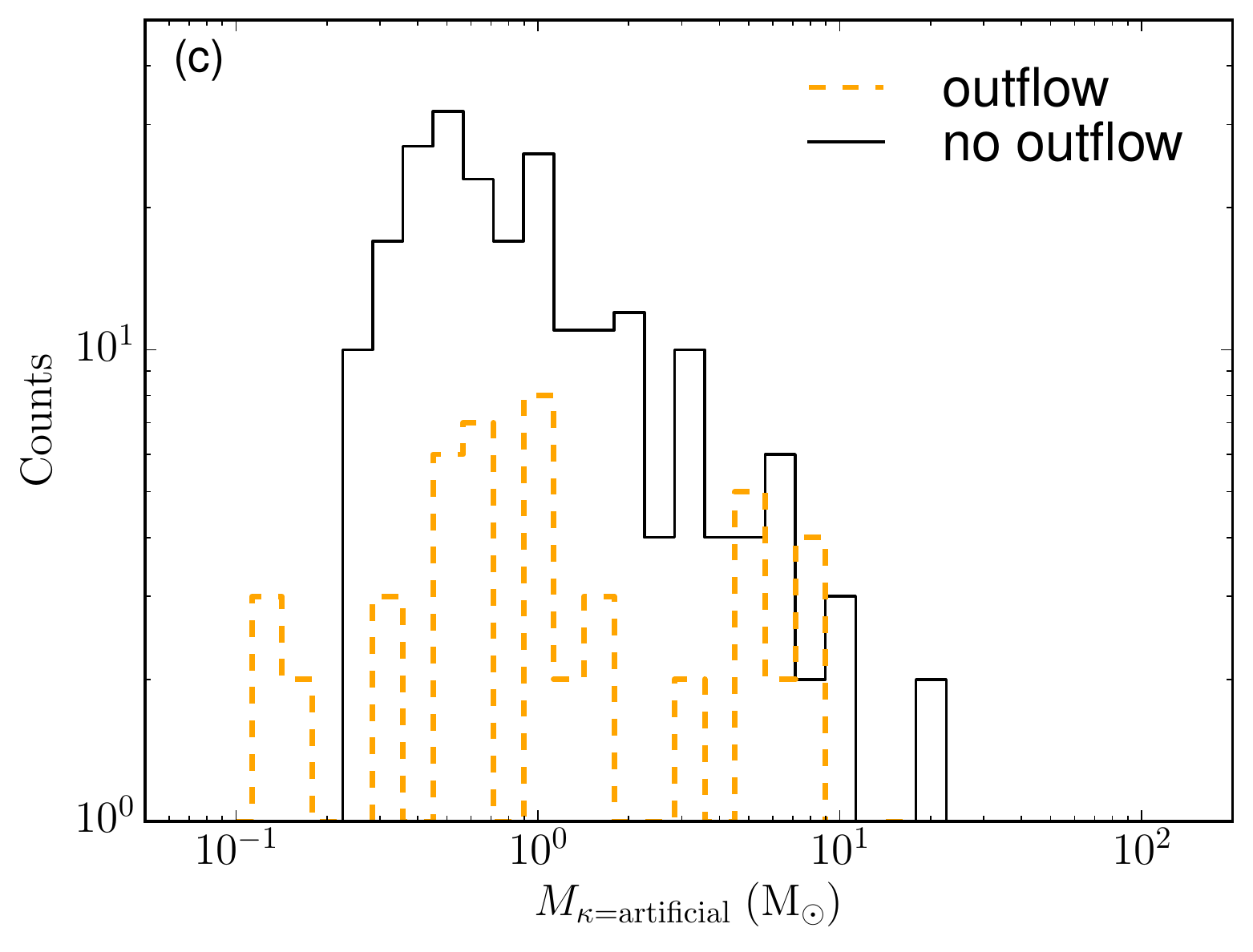}
\caption{
{\bf (a):} Gas temperatures for the
protostellar cores and the starless cores
for the {\it astrograph} sample.
{\bf (b):} Core mass histograms with
artificial core temperatures. Here, 
the starless cores are given a 15 K temperature
while the protostellar cores are set to 50 K.
{\bf (c):} Core mass histograms with 
an artificial opacity (for protostellar cores)
that is 4 times higher than the fiducial value 
(simply reducing protostellar core masses by
a factor of 4).
\label{fig:tkinarti}}
\end{figure}

There are several possible biases which must be carefully explored, especially for those parameters that could be systematically different between protostellar and starless cores. 
First, the  size of the synthesized beam of the VLA NH$_3$ data (6.5$\arcsec\times$3.6\arcsec) is larger than the core size ($\la$3\arcsec). 
Energy from the protostellar accretion and outflow can heat the core, resulting in an increasing temperature gradient toward the center of the core. Not properly resolving the core may  result in an underestimation of the core temperature, which in part results in an overestimation  of the core mass. 
Second, the NH$_3$ (1,1) and (2,2) lines in \citet{2018RNAAS...2...52W} probably cannot trace temperatures higher than 30 K.
These could make the protostellar cores ``appear'' to be more massive than what they really are. In Figure \ref{fig:tkinarti}(a), we show the gas temperatures for the two core populations in the {\it astrograph} sample. We can see that their temperatures are indistinguishable. Thus, the protostellar core temperatures could be underestimated.

However, the dust temperature map based on Herschel data \citep{2017ApJ...840...22L} showed that the majority of the Dragon IRDC is below 20 K, very similar to the gas temperature traced by the NH$_3$ measurement (W18). But note that the spatial resolution of the dust temperature map is 10\arcsec, i.e., a factor of $\sim$3 larger than the ALMA 1.3 mm continuum cores. So local enhancements of temperature by the protostars may not be visible. However, at least at the scale of VLA synthesized beam (6.5$\arcsec\times$3.6\arcsec), the enhancement of temperature is not seen. 

On the other hand, \citet{2015ApJ...804..141Z} and \citet{2017ApJ...834..193K} have shown the presence of significant N$_2$D$^+$ in many of the protostellar cores. For example, the most massive protostellar core is the C1-Sa core, which was studied in detail by \citet{2018ApJ...867...94K}. The authors showed that a significant amount of N$_2$D$^+$ remains in the core. The molecular ion implies low gas temperature $\la$15 K \citep{2015ApJ...804...98K}. These results suggest that the bulk of the mass of the protostellar cores are not significantly heated, and our assumptions with regards to the core temperature do not result in a significant bias. 

What temperature is required if we wish for the protostellar core mass to be indistinguishable from the starless core? In Figure \ref{fig:tkinarti}(b), we show that by arbitrarily giving the starless cores a lower temperature (15 K), and the protostellar cores a higher temperature (50 K) the mass distribution of these two core populations are basically indistinguishable. A Mann-Whitney U test results in a p-value of 0.19, which indicates that the null hypothesis that the protostellar cores are equally or less massive than the starless cores cannot be rejected with a high confidence. 
Note, with these arbitrary temperatures, the maximum starless core mass is 26 M$_\odot$; and the maximum protostellar core mass is only 9 M$_\odot$. Further reducing the starless core temperature will artificially ``make'' many massive starless cores. Further increasing the protostellar core temperature will make all of them low-mass cores. 


\subsection{Potential Biases from Opacity}

Another possible source of bias is our assumption with respect to the dust opacity. In the core mass calculation, we have adopted a constant dust opacity from the moderately coagulated thin ice mantle model from OH94. If the protostellar cores have systematically larger opacities (e.g., due to further dust coagulation), then their masses are overestimated. In turn, the protostellar cores might not be more massive. 

However, the opacity we choose is already for dust after a coagulation time of 10$^5$ yr (at a density of $n_H=10^6$ cm$^{-3}$). For ice mantle models at a higher density ($n_H=10^8$ cm$^{-3}$, OH94), the maximum opacity can reach 1.11 cm$^2$ g$^{-1}$, i.e., a factor of 1.2 higher. 
Even if we apply a factor of $\sim$2 higher opacity (compared to the fiducial value) for the protostellar cores,
the Mann-Whitney U test still rejects the null hypothesis that protostellar cores are equally or less massive than starless cores with $>$99.99\% confidence for both samples.

The opacity of dust particles without ice mantles can be larger by a factor of $\sim$6 compared to the fiducial value.
However, the particles in dark, cold cores should have a considerable amount of ice, which is supported by evidence of depletion \citep[e.g.,][]{2002A&A...389L...6B,2018ApJ...856..141T}. 

What opacity is required in order for the protostellar core masses to be indistinguishable from those of the starless core? Figure \ref{fig:tkinarti}(c) shows new core mass histograms. Here, the protostellar core masses are computed with an opacity 4 times higher than the fiducial value
(simply reducing protostellar core masses by a factor of 4). This time the null hypotheis that the protostellar cores are equally or less massive than the starless cores is only rejected with a confidence of 82.6\%. The factor of 4 can be achieved if dust particles in protostellar cores have no ice mantles during the coagulation. However, it is hard to imagine that dust in protostellar cores evolves differently, especially before the cores become protostellar.

Another possibility is that the protostellar cores have evolved much longer than the starless cores, so dust in the protostellar cores has more time to coagulate (e.g., much longer than 10$^5$ yr). That would indicate these protostellar cores have had a long starless phase. If this scenario were to be correct, then it would imply a fraction of the starless cores should also have underestimated dust opacity.

\subsection{Potential Biases from the Combination of Temperature and Opacity}


\begin{figure}[htb!]
\centering
\epsscale{1.}
\plotone{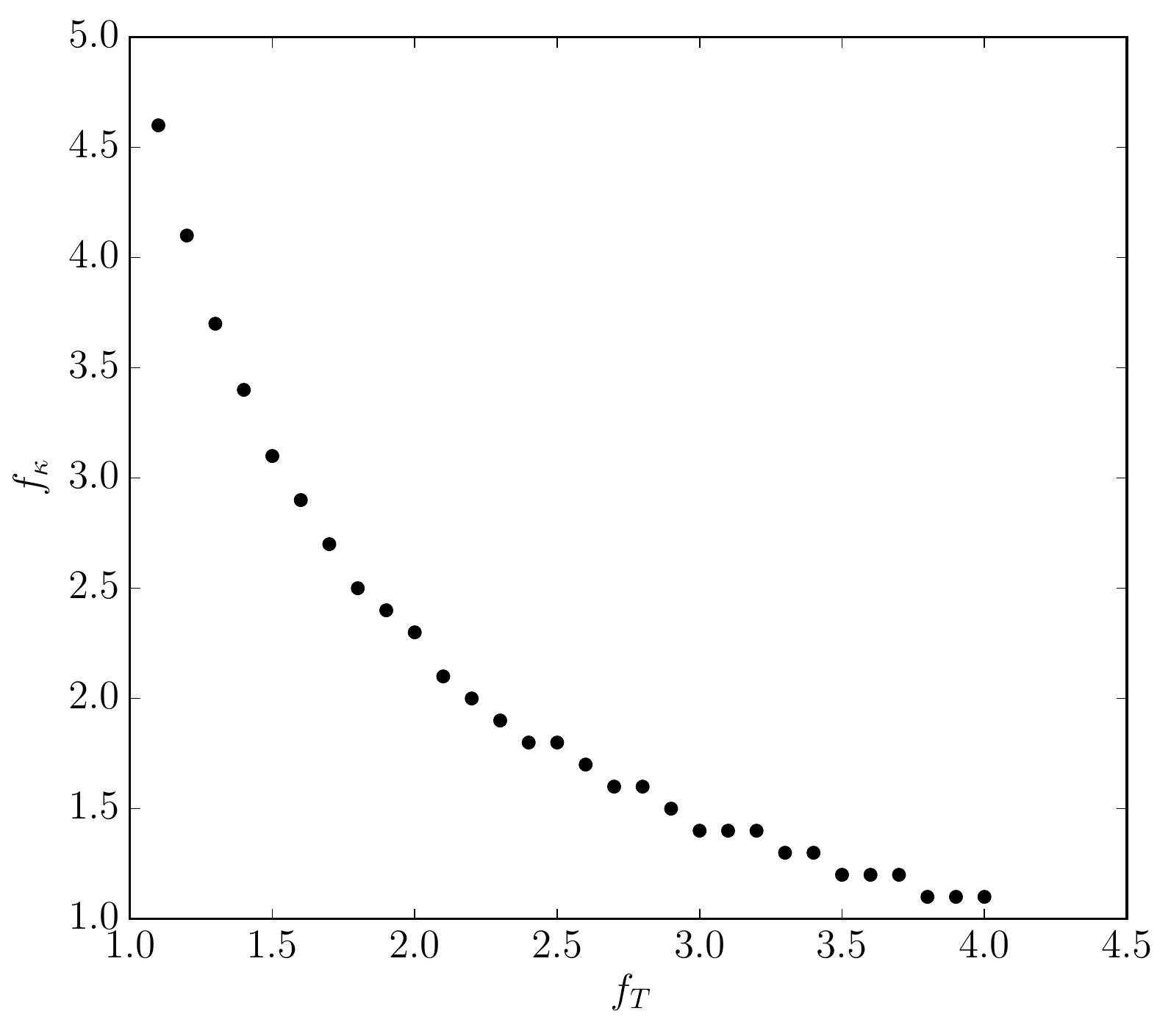}
\caption{
Factors used to manually make
protostellar core masses indistinguishable 
from starless core masses.
The x-axis is the factor for increasing 
the protostellar core temperature ($f_T$).
The y-axis is the factor for increasing 
the protostellar core opacity ($f_\kappa$).
\label{fig:artiTk}}
\end{figure}


In Figure \ref{fig:artiTk}, we investigate
what combinations of arbitrarily increased 
protostellar core temperature and opacity
make the protostellar core mass indistinguishable
from the starless core mass. Specifically,
we increase the protostellar core temperature
and opacity by $f_T$ and $f_\kappa$, respectively.
With a $f_T$ value, we iteratively find the
corresponding $f_\kappa$ that results in 
a Mann-Whitney U test p-value of 0.5, i.e.,
that the mass of one core population is
higher or lower than the that of the other population is rejected with a 50\% confidence.

A combination of underestimated protostellar core dust temperature and opacity that overestimate the protostellar core mass by a factor of $\sim$4-5 could in principle account for the seemingly more massive protostellar cores.
As shown in Figure \ref{fig:artiTk}, if the 
protostellar core temperature is accurate
($f_T$ close to 1), then the opacity has to
be underestimated by a factor of $f_\kappa\sim$5
so that the protostellar cores have similar 
masses as the starless cores. Meanwhile, if the
opacity is accurate (close to 1), the 
temperature has to be underestimated by 
$f_T\sim$4 for the masses to be similar between
the two core populations. If both the temperature
and opacity are underestimated for protostellar 
cores, a combined underestimation factor 
of $f_T f_\kappa\sim$4-5 is needed to explain
the mass difference between the two core 
populations. In the future, 
ALMA may be able to better constrain the dust
opacity through a multi-wavelength continuum 
survey; and planned facilities like the ngVLA could give
better estimates of core temperatures using NH$_3$ observations.

\section{Discussion}\label{sec:disc}

In our sample of cores in the Dragon IRDC, the protostellar cores are statistically more massive and larger than the starless cores. We interpret this difference as evidence that cores grow in mass and volume during their evolution from the starless stage to the protostellar stage. In this simple picture, cores are constantly formed. Cores that form earlier collapse and form protostars (with outflows). Cores that form later are starless and may or may not become protostellar. Once the cores do become protostellar, they continue to grow in the IRDC and appear more massive
\citep[see related discussions in][]{2020arXiv200607325C}. With the ALMA image, which only provides a snapshot of the whole process, we see outflows more likely associated with more massive cores. 

However, one could argue against this by saying that more massive cores evolve faster. In this alternative scenario,  all cores roughly emerge at the same time (due to, e.g., fragmentation), and those more massive produce protostars earlier. Consequently, the cores we identify as protostellar are not more massive due to core growth but their mass is a result of the initial fragmentation process. We can evaluate this alternative picture by comparing the free-fall timescales (which depends on the core densities) for the two types of cores. As we have shown in Figure \ref{fig:dens}, the protostellar and starless cores have statistically indistinguishable average density distributions. Consequently, it is unlikely that the protostellar cores evolve faster than the starless cores. The fact that some cores host outflows imply that these have been able to collapse and form protostars before the rest of the cores.

The core growth idea is consistent with the filament-core accretion picture proposed by \citet{2019ApJ...874..104K}, where they suggested that filaments feed the embedded cores and protostars with additional mass. A recent numerical work by \citet{2020ApJ...900...82P} showed that star formation begins with low-to-intermediate mass cores that are gravitationally unstable. In this scenario of (massive) star formation, which the authors refer to as the ``inertial-inflow'' model, the final mass of a massive star is much larger than the initial core mass, and the accretion time for the massive star formation is a few times the initial free-fall time. If filament-core accretion/inertial-inflow indeed take place in  IRDCs, our finding of core mass growth will raise questions for the relationship between CMF and the stellar initial mass function (IMF) because the CMF will not be static. The CMF-IMF relation would depend on the mass infall rate and timescale, the core-to-star efficiency, and the binarity fraction. In addition, the filament-core accretion process must end at some point, most likely due to feedback from massive stars, and this could also be important in determining the final stellar mass \citep{2018A&A...616A.101K}.  

\section{Conclusion}\label{sec:conc}

In summary, we investigate the core mass distribution between protostellar cores (cores with molecular outflows) and starless cores (no infrared sources, no CO or SiO outflows) in the Dragon IRDC. We find that the protostellar cores are statistically more massive than the starless cores. Further analyses show that the protostellar cores have statistically larger sizes but similar densities as compared to starless cores. We suggest that the mass difference is caused by continuous core growth since their formation, unless the mean temperature and opacity are underestimated by a total factor of 4-5 for protostellar cores.

A potential scenario that may explain our results is that  cores grow by acquiring the inflowing material that is channeled by the filament in which the cores emerge. Depending on how much more mass a core gains during its lifetime, the core growth may resolve the issue of the lack of massive prestellar cores in massive star-forming regions, because massive stars begin with low-to-intermediate mass cores.

\acknowledgments 
We thank the anonymous referee for helpful comments.
We thank John Bieging, Ke Wang for fruitful discussions.
HGA acknowledges support from  NSF award AST-1714710.
This paper makes use of the following ALMA data:
ADS/JAO.ALMA\#2013.1.00183.S and ADS/JAO.ALMA\#2015.1.00183.S. 
ALMA is a partnership of ESO
(representing its member states), NSF (USA) and NINS (Japan), together
with NRC (Canada), NSC and ASIAA (Taiwan), and KASI (Republic of
Korea), in cooperation with the Republic of Chile.  The Joint ALMA
Observatory is operated by ESO, AUI/NRAO and NAOJ.  The National Radio
Astronomy Observatory is a facility of the National Science Foundation
operated under cooperative agreement by Associated Universities, Inc.

\software{Python \citep{python}, SciPy \citep{scipy}, Astropy \citep{Astropy-Collaboration13}, Numpy \citep{numpy}, Matplotlib \citep{matplotlib}, SAOImageDS9 \citep{2003ASPC..295..489J}}

\facility{ALMA, VLA, Effelsberg 100m}.


\appendix
\restartappendixnumbering 

\section{Core Kinematics}\label{app:kin}

We derive the core kinematics by fitting Gaussians to core spectral line profiles, including C$^{18}$O(2-1), N$_2$D$^+$(3-2), and DCO$^+$(3-2). The spectral line profiles are averaged over the projected core area in the spectral line cubes.
For each core, if it has an outflow (K19b), we fit a Gaussian (or multiple Gaussians) to the C$^{18}$O emission. If the core shows multiple components, we make integrated intensity maps for each component and compare them with the continuum image. We choose the associated velocity component based on  morphology matching. If the core has no outflow, i.e., it is starless, we search for emission features from any of the N$_2$D$^+$ and DCO$^+$ lines. The underlying assumption is that starless cores suffer from CO depletion due to freeze-out \citep{2005ApJ...619..379C}. So for cores with no outflows, the C$^{18}$O line profile may not trace the core kinematics as robust as the deuterated species.

Occasionally, a starless core only has C$^{18}$O emission (no N$_2$D$^+$ or DCO$^+$); or a protostellar core only has emission from N$_2$D$^+$ or DCO$^+$. In that case, we adopt the velocity from these lines only if it is between 75 to 85 km s$^{-1}$, and the morphology of the integrated intensity map matches the continuum. The former requirement is based on the observation that the cloud velocity on a large scale is at $\sim$ 79 km s$^{-1}$. The latter is to make sure (to our best effort) that the line emission is associated with the continuum.

\begin{figure*}[htb!]
\centering
\epsscale{1.1}
\plotone{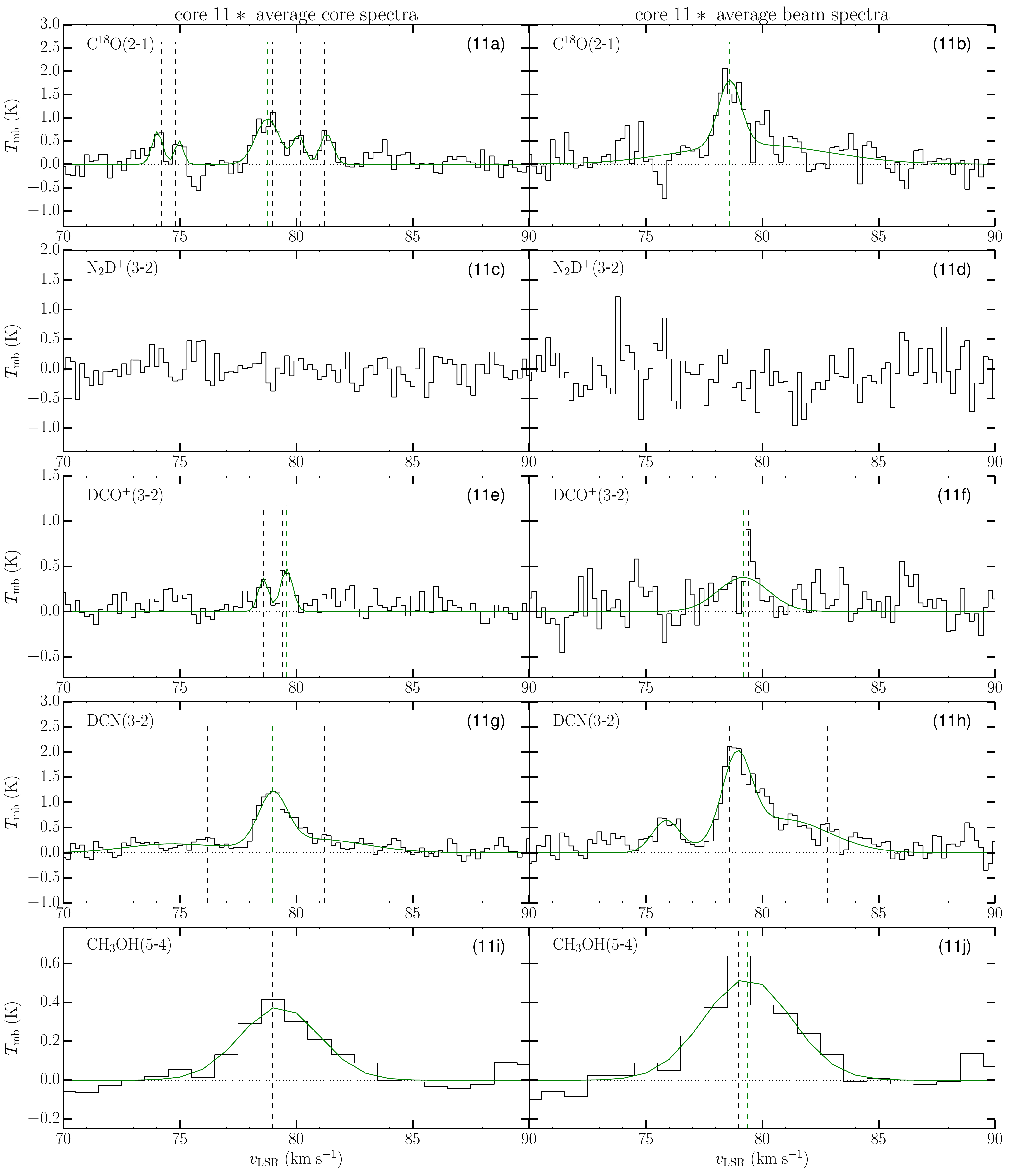}
\caption{
Example for a spectral line fitting for
core 11. The black step-curve shows the
raw spectrum. The green curve shows the
multi-Gaussian fitting to the spectrum.
The black dashed vertical line shows the
velocity of the initial guess (local peak)
of each Gaussian component. The green 
dashed line shows the peak velocity for
the strongest fitted Gaussian component. 
The left column shows averaged spectra 
within the core while the right shows 
those within the beam. The asterisk following
the core name in the title indicates the core
has outflows.
\label{fig:spec}}
\end{figure*}

Figure \ref{fig:spec} shows an example of the spectra for core 11. In each spectrum, if at least three adjacent channels have a signal-to-noise ratio greater than 2, the component is included in the initial guess for a Gaussian component. Then, a multi-Gaussian fitting is performed. The N$_2$D$^+$(3-2) line has multiple hyperfine components. However, the signal-to-noise in the data is such that
only its strongest central hyperfine component is detected, if any. Hence, the Gaussian fitting is a reasonable approximation. In total, only about ten cores use the N$_2$D$^+$ fitting result.

\begin{figure*}[htb!]
\centering
\epsscale{1.1}
\plotone{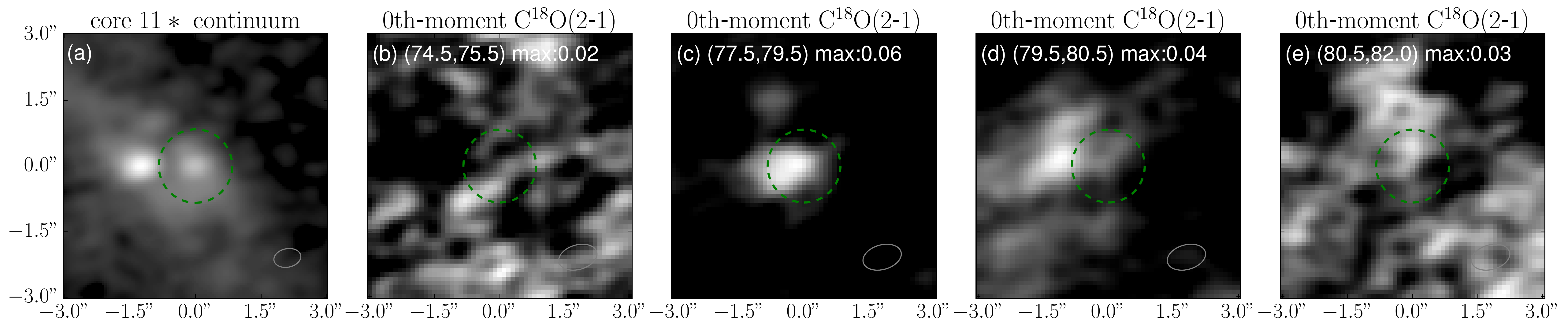}
\caption{
Comparison between the continuum image for 
core 11 and the integrated intensity maps
for all components between 75 km s$^{-1}$
and 85 km s$^{-1}$. The numbers in parentheses at the top of all panels, except for panel (a),
indicate the integration
velocity range in km s$^{-1}$. The label
``max'' gives the maximum intensity in the panel in units of Jy beam$^{-1}$ km s$^{-1}$.
\label{fig:mom0}}
\end{figure*}

Figure \ref{fig:mom0} shows a comparison between the continuum and the integrated intensity maps for core 11. Four line components are included for this comparison. The structure in panel (c) 
shows the best coherent structure around the core, so its fitting result is adopted. All 280 spectra figures and all 180 comparison figures (for those with line detection) are available at \url{https://doi.org/10.7910/DVN/OLRED4}.

Table \ref{tab:kin} includes the kinematic information for all 280 {\it astrograph} cores. The core names are listed in column (1). Columns (2) and (3) show the core coordinates in J2000 degrees. Column (4) lists the number of components detected in each line (C$^{18}$O,N$_2$D$^+$,DCO$^+$,DCN,CH$_3$OH). A dash sign right of the number indicates  a tentative detection (i.e., a low signal-to-noise component). 
Column (5) displays which line is adopted and how many components to check in the integrated intensity maps (the number in the brackets). The number of components to check 
can be smaller than the total number of components in the line fitting because we do not check low-confidence components (but still report them). 
Columns (6) and (7) list the LSR velocities and dispersion  from the fit to the spectrum averaged over projected core area. Columns (8) and (9) are the velocity of the peak emission and dispersion from a fit to the spectrum averaged within a beam. Column (10) is the total velocity dispersion, where the line thermal component is subtracted, and the sound speed is added back (using the gas temperature from W18 and assuming a mean molecular weight per free particle of 2.37).
Column (11) is the total velocity dispersion based on the NH$_3$ dispersion given by W18. 

\newpage

\startlongtable
\begin{deluxetable*}{ccccccccccc}
\tabletypesize{\scriptsize}
\tablewidth{0pt} 
\tablecaption{Core Kinematics \label{tab:kin}}
\tablehead{
\colhead{Core} & \colhead{RA}& \colhead{DEC} & \colhead{Detection} & \colhead{Fitting} & \colhead{$v_{\rm lsr,c}$} & \colhead{$\sigma_{\rm fit,c}$} & \colhead{$v_{\rm lsr,b}$} & \colhead{$\sigma_{\rm fit,b}$} & \colhead{$\sigma_{\rm tot}$} & \colhead{$\sigma_{\rm tot,NH_3}$}\\
\colhead{} & \colhead{deg}& \colhead{deg} & \colhead{} & \colhead{line} & \colhead{km s$^{-1}$} & \colhead{km s$^{-1}$} & \colhead{km s$^{-1}$} & \colhead{km s$^{-1}$} & \colhead{km s$^{-1}$} & \colhead{km s$^{-1}$}
} 
\colnumbers
\startdata 
1 & 280.69353 & -4.07084 & 3,1,1,1,1 & C$^{18}$O [3] & 78.97(0.03) & 0.30(0.03) & 78.94(0.02) & 0.39(0.02) & 0.36(0.03) & 0.83(0.01) \\
2 & 280.71076 & -4.05449 & 2,0,0,1,1 & C$^{18}$O [2] & 79.22(0.07) & 1.17(0.07) & 78.98(0.04) & 0.66(0.04) & 1.19(0.07) & 0.87(0.01) \\
3 & 280.71328 & -4.05201 & 2,1,1,1,0 & C$^{18}$O [2] & 78.99(0.03) & 0.60(0.03) & 79.10(0.03) & 0.51(0.03) & 0.64(0.03) & 0.82(0.01) \\
4 & 280.72522 & -4.04092 & 1,0,1,1,1 & C$^{18}$O [1] & 79.64(0.05) & 0.59(0.05) & 79.63(0.03) & 0.45(0.03) & 0.64(0.05) & 1.06(0.01) \\
5 & 280.71186 & -4.05317 & 1,0,1,3,1 & C$^{18}$O [1] & 79.58(0.05) & 0.53(0.05) & 79.53(0.07) & 0.73(0.07) & 0.58(0.05) & 1.09(0.01) \\
6 & 280.70758 & -4.05705 & 1,1,1-,1,0 & N$_2$D$^+$ [1] & 79.38(0.05) & 0.33(0.05) & 79.33(0.06) & 0.31(0.06) & 0.39(0.04) & 0.99(0.01) \\
7 & 280.70419 & -4.03817 & 0,0,0,0,0 & - & - & - & - & - & - & 0.64(0.00) \\
8 & 280.70950 & -4.03319 & 1,0,1,2,1- & C$^{18}$O [1] & 78.66(0.03) & 0.86(0.03) & 78.84(0.04) & 0.92(0.04) & 0.89(0.03) & 0.75(0.01) \\
9 & 280.72832 & -4.03704 & 1,0,0,1,1- & C$^{18}$O [1] & 80.41(0.03) & 0.30(0.03) & - & - & 0.37(0.03) & 0.77(0.01) \\
10 & 280.72574 & -4.04240 & 1,0,1,1,1- & C$^{18}$O [1] & 79.89(0.07) & 0.74(0.07) & 79.70(0.06) & 0.71(0.06) & 0.77(0.07) & 0.77(0.01) \\
11 & 280.71151 & -4.05317 & 5,0,2,3,1 & C$^{18}$O [4] & 78.77(0.07) & 0.51(0.07) & 78.60(0.07) & 0.49(0.08) & 0.56(0.06) & 1.09(0.01) \\
12 & 280.70946 & -4.05559 & 2-,1,1,1,1- & C$^{18}$O [1] & 78.85(0.07) & 0.59(0.06) & 78.73(0.06) & 0.72(0.06) & 0.63(0.06) & 0.92(0.01) \\
13 & 280.72497 & -4.04333 & 1,0,1,0,1 & C$^{18}$O [1] & 79.05(0.07) & 0.60(0.07) & 78.83(0.05) & 0.56(0.05) & 0.62(0.07) & 0.71(0.01) \\
14 & 280.72921 & -4.03662 & 0,0,0,0,1 & - & - & - & - & - & - & 0.79(0.01) \\
15 & 280.70303 & -4.03926 & 2,0,0,3-,1 & C$^{18}$O [2] & 82.66(0.06) & 0.50(0.06) & 82.60(0.06) & 0.53(0.06) & 0.55(0.05) & 0.89(0.01) \\
16 & 280.70544 & -4.04009 & 3-,0,1,0,0 & DCO$^+$ [1] & 79.92(0.03) & 0.27(0.03) & 79.69(0.12) & 0.52(0.12) & 0.32(0.03) & 0.70(0.01) \\
17 & 280.72752 & -4.03933 & 1,0,1,1,1 & DCO$^+$ [1] & 79.42(0.02) & 0.39(0.02) & 79.45(0.03) & 0.43(0.03) & 0.46(0.02) & 0.84(0.01) \\
18 & 280.71357 & -4.05177 & 1,0,1,0,0 & DCO$^+$ [1] & 79.62(0.06) & 0.46(0.06) & - & - & 0.51(0.05) & 0.86(0.01) \\
19 & 280.73045 & -4.03949 & 2,0,0,1,1- & C$^{18}$O [1] & 78.86(0.05) & 0.52(0.05) & 78.96(0.04) & 0.44(0.04) & 0.56(0.05) & 0.77(0.01) \\
20 & 280.70918 & -4.05562 & 1,0,1,1,1 & DCO$^+$ [1] & 78.41(0.10) & 0.49(0.10) & 78.53(0.11) & 0.59(0.11) & 0.54(0.09) & 0.92(0.01) \\
21 & 280.70451 & -4.03592 & 1,0,0,1-,0 & C$^{18}$O [1] & 79.65(0.03) & 0.36(0.03) & - & - & 0.41(0.03) & 0.78(0.01) \\
22 & 280.73035 & -4.02063 & 1,0,1,1,1- & C$^{18}$O [1] & 78.25(0.03) & 0.57(0.03) & 78.16(0.03) & 0.60(0.03) & 0.60(0.03) & 0.67(0.01) \\
23 & 280.70286 & -4.03902 & 1,0,1-,0,1- & C$^{18}$O [1] & 82.63(0.05) & 0.41(0.05) & 82.60(0.04) & 0.42(0.04) & 0.47(0.04) & 0.91(0.01) \\
24 & 280.72403 & -4.03774 & 0,0,0,0,0 & - & - & - & - & - & - & 0.70(0.01) \\
25 & 280.70409 & -4.04170 & 2-,0,0,0,0 & C$^{18}$O [1] & 81.20(0.03) & 0.16(0.03) & - & - & 0.28(0.02) & 0.73(0.00) \\
26 & 280.71122 & -4.05426 & 1,0,0,0,0 & C$^{18}$O [1] & - & - & - & - & - & 0.97(0.01) \\
27 & 280.72400 & -4.03801 & 3-,0,0,0,1- & C$^{18}$O [2] & 79.82(0.11) & 0.37(0.11) & 79.44(0.17) & 0.68(0.17) & 0.43(0.10) & 0.70(0.01) \\
28 & 280.70969 & -4.05597 & 3,1-,2,1,1 & DCO$^+$ [2] & 79.24(0.09) & 0.51(0.10) & - & - & 0.56(0.09) & 0.84(0.00) \\
29 & 280.70544 & -4.04104 & 1,0,0,1-,0 & C$^{18}$O [1] & - & - & - & - & - & 0.68(0.00) \\
30 & 280.72710 & -4.03934 & 2-,0,1,0,1- & DCO$^+$ [1] & 79.32(0.03) & 0.30(0.03) & 79.28(0.03) & 0.31(0.03) & 0.39(0.02) & 0.84(0.01) \\
31 & 280.70441 & -4.04070 & 2,0,0,0,0 & C$^{18}$O [2] & - & - & - & - & - & 0.77(0.00) \\
32 & 280.70800 & -4.05699 & 3,1,1,0,1- & DCO$^+$ [1] & 79.29(0.12) & 0.49(0.13) & 79.44(0.16) & 0.87(0.17) & 0.53(0.12) & 0.99(0.01) \\
33 & 280.69402 & -4.07155 & 1,0,2-,0,0 & C$^{18}$O [1] & 81.18(0.04) & 0.53(0.04) & 81.09(0.06) & 0.58(0.06) & 0.56(0.04) & 0.87(0.01) \\
34 & 280.71328 & -4.05142 & 1,0,1,1,1 & DCO$^+$ [1] & 78.87(0.07) & 0.48(0.07) & 78.99(0.07) & 0.43(0.07) & 0.53(0.06) & 0.86(0.01) \\
35 & 280.70992 & -4.05583 & 2,0,1,2,1- & C$^{18}$O [2] & 78.58(0.03) & 0.55(0.03) & 78.61(0.02) & 0.52(0.03) & 0.59(0.03) & 0.91(0.01) \\
36 & 280.71303 & -4.05163 & 1,0,2-,1,1 & C$^{18}$O [1] & 78.93(0.02) & 0.42(0.02) & 78.96(0.02) & 0.41(0.02) & 0.47(0.02) & 0.87(0.01) \\
37 & 280.72021 & -4.04612 & 2-,0,0,0,0 & C$^{18}$O [1] & 78.91(0.07) & 0.50(0.07) & 78.89(0.07) & 0.56(0.07) & 0.54(0.07) & 0.65(0.01) \\
38 & 280.70465 & -4.04199 & 2,0,0,1,1 & C$^{18}$O [2] & 79.43(0.05) & 0.24(0.05) & 79.41(0.04) & 0.23(0.04) & 0.33(0.04) & 0.80(0.01) \\
39 & 280.71168 & -4.05516 & 2,1,0,0,0 & N$_2$D$^+$ [1] & 78.67(0.10) & 0.48(0.10) & 78.68(0.10) & 0.32(0.10) & 0.53(0.09) & 1.15(0.01) \\
40 & 280.72511 & -4.04151 & 2,0,1,0,1 & C$^{18}$O [2] & 78.89(0.07) & 0.61(0.07) & 78.92(0.05) & 0.50(0.05) & 0.65(0.07) & 0.94(0.01) \\
41 & 280.68586 & -4.07262 & 3-,0,0,0,0 & C$^{18}$O [2] & 78.35(0.14) & 0.87(0.12) & 78.25(0.11) & 0.89(0.11) & 0.89(0.12) & 0.76(0.01) \\
42 & 280.72510 & -4.04133 & 1,0,0,1,1 & C$^{18}$O [1] & - & - & 80.99(0.08) & 0.35(0.08) & - & 1.06(0.01) \\
43 & 280.71131 & -4.05386 & 0,0,0,0,0 & - & - & - & - & - & - & 1.03(0.01) \\
44 & 280.72117 & -4.04202 & 1,0,0,1,0 & C$^{18}$O [1] & 80.03(0.04) & 0.36(0.04) & 80.05(0.05) & 0.39(0.05) & 0.41(0.04) & 0.69(0.01) \\
45 & 280.70717 & -4.04238 & 1,0,0,0,2- & C$^{18}$O [1] & 79.65(0.07) & 0.50(0.07) & 79.61(0.08) & 0.54(0.08) & 0.54(0.07) & 0.65(0.01) \\
46 & 280.72534 & -4.03817 & 1-,0,0,0,1- & - & - & - & - & - & - & 0.96(0.01) \\
47 & 280.71892 & -4.04391 & 1,0,1,0,0 & DCO$^+$ [1] & 80.81(0.07) & 0.52(0.07) & 80.76(0.06) & 0.45(0.06) & 0.56(0.06) & 0.94(0.01) \\
48 & 280.70421 & -4.03679 & 1,0,1-,0,0 & C$^{18}$O [1] & - & - & - & - & - & 0.78(0.01) \\
49 & 280.72399 & -4.04140 & 1,0,1,0,0 & C$^{18}$O [1] & 79.38(0.05) & 0.30(0.05) & 79.37(0.06) & 0.31(0.06) & 0.36(0.04) & 0.83(0.01) \\
50 & 280.71137 & -4.05470 & 3,0,0,0,1 & C$^{18}$O [3] & 77.92(0.07) & 0.54(0.07) & 77.88(0.05) & 0.45(0.05) & 0.58(0.07) & 0.97(0.01) \\
51 & 280.71929 & -4.04615 & 3-,0,0,0,0 & C$^{18}$O [2] & 79.17(0.03) & 0.60(0.03) & 79.18(0.03) & 0.58(0.03) & 0.64(0.03) & 0.85(0.01) \\
52 & 280.70231 & -4.03840 & 0,0,2-,0,0 & - & - & - & - & - & - & 0.85(0.01) \\
53 & 280.71035 & -4.05629 & 3,0,1,0,1- & DCO$^+$ [1] & 78.07(0.10) & 0.42(0.10) & 78.04(0.10) & 0.38(0.10) & 0.47(0.09) & 0.87(0.01) \\
54 & 280.69561 & -4.06440 & 1,0,0,1,0 & C$^{18}$O [1] & 80.69(0.13) & 0.58(0.13) & 80.92(0.07) & 0.28(0.07) & 0.61(0.12) & 1.04(0.01) \\
55 & 280.70946 & -4.03281 & 1,0,0,0,0 & C$^{18}$O [1] & 79.25(0.18) & 1.03(0.18) & 79.33(0.20) & 0.96(0.20) & 1.05(0.18) & 0.73(0.01) \\
56 & 280.70218 & -4.03879 & 0,1,0,0,0 & N$_2$D$^+$ [1] & 80.14(0.08) & 0.25(0.08) & - & - & 0.34(0.06) & 0.89(0.01) \\
57 & 280.71216 & -4.05279 & 0,0,0,0,0 & - & - & - &  &  & - & 0.88(0.01) \\
58 & 280.72112 & -4.04554 & 1,0,0,0,0 & C$^{18}$O [1] & 79.28(0.06) & 0.29(0.06) & 79.27(0.05) & 0.28(0.05) & 0.36(0.05) & 0.73(0.01) \\
59 & 280.72440 & -4.04402 & 1,0,0,0,1- & C$^{18}$O [1] & 79.19(0.02) & 0.30(0.02) & 79.18(0.02) & 0.31(0.02) & 0.36(0.02) & 0.72(0.01) \\
60 & 280.70417 & -4.03576 & 1,0,0,1-,1- & C$^{18}$O [1] & 79.41(0.24) & 1.31(0.24) & - & - & 1.33(0.24) & 0.78(0.01) \\
61 & 280.71179 & -4.05291 & 1,0,0,1-,1- & C$^{18}$O [1] & 79.61(0.07) & 0.61(0.07) & 79.58(0.06) & 0.51(0.06) & 0.65(0.07) & 1.01(0.01) \\
62 & 280.71901 & -4.04451 & 2,0,1,1-,0 & C$^{18}$O [2] & 81.55(0.17) & 0.37(0.17) & - & - & 0.43(0.15) & 0.97(0.01) \\
63 & 280.72522 & -4.04376 & 1,0,1,1,0 & C$^{18}$O [1] & 78.53(0.03) & 0.36(0.03) & 78.55(0.03) & 0.37(0.03) & 0.41(0.03) & 0.68(0.01) \\
64 & 280.70992 & -4.05622 & 1,0,1,1,1 & DCO$^+$ [1] & 78.23(0.06) & 0.64(0.06) & 78.28(0.07) & 0.66(0.07) & 0.68(0.06) & 0.86(0.01) \\
65 & 280.70829 & -4.03367 & 1,0,0,0,0 & C$^{18}$O [1] & 79.23(0.05) & 0.50(0.05) & 79.24(0.05) & 0.51(0.05) & 0.54(0.05) & 0.66(0.01) \\
66 & 280.70782 & -4.03937 & 0,0,0,0,0 & - & - & - & - & - & - & 0.77(0.01) \\
67 & 280.72638 & -4.04219 & 0,1-,0,0,0 & - & - & - & - & - & - & 0.64(0.01) \\
68 & 280.70306 & -4.03441 & 1,1,0,0,0 & N$_2$D$^+$ [1] & 81.04(0.14) & 0.46(0.14) & - & - & 0.51(0.13) & 0.66(0.01) \\
69 & 280.72568 & -4.04305 & 0,0,0,0,0 & - & - & - & - & - & - & 0.76(0.01) \\
70 & 280.70712 & -4.04119 & 0,0,0,0,0 & - & - & - & - & - & - & 0.65(0.01) \\
71 & 280.72375 & -4.03840 & 1-,0,1-,0,0 & - & - & - & - & - & - & 0.74(0.01) \\
72 & 280.70239 & -4.03888 & 1-,0,1,0,1- & DCO$^+$ [1] & 81.25(0.21) & 0.86(0.21) & 80.82(0.15) & 0.57(0.15) & 0.89(0.20) & 0.89(0.01) \\
73 & 280.72563 & -4.04056 & 0,0,0,0,0 & - & - & - & - & - & - & 0.87(0.01) \\
74 & 280.70185 & -4.06808 & 1,0,0,0,0 & C$^{18}$O [1] & 81.91(0.10) & 0.34(0.10) & - & - & 0.37(0.09) & 0.90(0.02) \\
75 & 280.70597 & -4.06677 & 1,0,0,0,0 & C$^{18}$O [1] & 78.80(0.05) & 0.48(0.05) & 78.80(0.05) & 0.49(0.05) & 0.52(0.05) & 0.85(0.01) \\
76 & 280.72152 & -4.04641 & 1,0,0,1,0 & C$^{18}$O [1] & 79.20(0.07) & 0.43(0.07) & 79.25(0.09) & 0.49(0.09) & 0.47(0.06) & 0.72(0.01) \\
77 & 280.70388 & -4.03895 & 0,0,0,0,1- & - & - & - & - & - & - & 0.83(0.01) \\
78 & 280.68737 & -4.07448 & 2,0,1-,0,0 & C$^{18}$O [2] & 78.09(0.04) & 0.55(0.04) & 78.13(0.04) & 0.55(0.04) & 0.58(0.04) & 0.87(0.02) \\
79 & 280.70668 & -4.04230 & 0,0,0,0,0 & - & - & - & - & - & - & 0.67(0.01) \\
80 & 280.71143 & -4.05498 & 2-,0,0,0,0 & C$^{18}$O [1] & 80.24(0.09) & 0.49(0.09) & 80.18(0.12) & 0.55(0.12) & 0.54(0.08) & 1.15(0.01) \\
81 & 280.71971 & -4.04866 & 1,0,0,0,0 & C$^{18}$O [1] & 78.33(0.05) & 0.55(0.05) & 78.36(0.05) & 0.56(0.05) & 0.59(0.05) & 0.78(0.01) \\
82 & 280.68691 & -4.06826 & 1,0,1,0,0 & C$^{18}$O [1] & 79.49(0.05) & 0.25(0.05) & 80.01(0.18) & 0.79(0.18) & 0.32(0.04) & 0.66(0.01) \\
83 & 280.70969 & -4.03233 & 0,0,0,0,0 & - & - & - & - & - & - & 0.72(0.01) \\
84 & 280.68766 & -4.07442 & 3,0,1,0,0 & C$^{18}$O [3] & 78.05(0.02) & 0.47(0.02) & 78.07(0.02) & 0.44(0.02) & 0.51(0.02) & 0.81(0.02) \\
85 & 280.71056 & -4.05499 & 4-,0,1-,1-,1- & C$^{18}$O [3] & 80.29(0.10) & 0.60(0.10) & 79.77(0.05) & 0.20(0.05) & 0.64(0.09) & 0.87(0.01) \\
86 & 280.69555 & -4.06926 & 1,0,0,0,0 & C$^{18}$O [1] & - & - & - & - & - & 0.84(0.01) \\
87 & 280.71108 & -4.05508 & 1,0,0,0,0 & C$^{18}$O [1] & - & - & - & - & - & 0.91(0.01) \\
88 & 280.69553 & -4.06959 & 0,0,0,0,0 & - & - & - & - & - & - & 0.84(0.01) \\
89 & 280.71253 & -4.05220 & 1,0,0,0,0 & C$^{18}$O [1] & 76.97(0.16) & 0.58(0.16) & - & - & 0.62(0.15) & 0.90(0.01) \\
90 & 280.72425 & -4.04386 & 1,0,1-,0,0 & C$^{18}$O [1] & 79.20(0.05) & 0.32(0.05) & 79.18(0.04) & 0.30(0.04)  & 0.37(0.04) & 0.72(0.01) \\
91 & 280.71367 & -4.05241 & 2,1,1-,0,0 & N$_2$D$^+$ [1] & 78.58(0.09) & 0.40(0.09) & 78.60(0.14) & 0.47(0.14) & 0.44(0.08) & 0.80(0.01) \\
92 & 280.71042 & -4.05531 & 1,0,0,0,0 & C$^{18}$O [1] & 80.41(0.03) & 0.26(0.03) & 80.44(0.04) & 0.26(0.04) & 0.34(0.02) & 0.90(0.01) \\
93 & 280.70967 & -4.05674 & 2,0,1-,1-,0 & C$^{18}$O [1] & 79.21(0.05) & 0.33(0.05) & 79.16(0.05) & 0.31(0.05) & 0.39(0.04) & 0.81(0.01) \\
94 & 280.72829 & -4.01469 & 1,0,0,1-,1- & C$^{18}$O [1] & 79.80(0.18) & 0.90(0.18) & 79.87(0.17) & 0.80(0.17) & 0.92(0.18) & 0.65(0.01) \\
95 & 280.71890 & -4.04509 & 1,1-,1,0,0 & DCO$^+$ [1] & 79.47(0.04) & 0.22(0.04) & 79.44(0.04) & 0.23(0.04) & 0.31(0.03) & 0.95(0.01) \\
96 & 280.69356 & -4.07159 & 2,0,0,0,0 & C$^{18}$O [2] & 81.47(0.09) & 0.60(0.09) & 81.45(0.09) & 0.53(0.09) & 0.63(0.09) & 0.90(0.01) \\
97 & 280.70600 & -4.04040 & 0,0,1,0,0 & DCO$^+$ [1] & 80.04(0.04) & 0.30(0.04) & 80.05(0.04) & 0.32(0.04) & 0.36(0.03) & 0.68(0.00) \\
98 & 280.72910 & -4.03695 & 0,0,0,0,0 & - & - & - & - & - & - & 0.80(0.01) \\
99 & 280.72499 & -4.04359 & 0,1,0,0,0 & N$_2$D$^+$ [1] & - & - & 79.42(0.16) & 0.50(0.16) & - & 0.71(0.01) \\
100 & 280.71027 & -4.05651 & 2,0,1,0,1- & DCO$^+$ [1] & 78.94(0.08) & 0.32(0.08) & 78.93(0.08) & 0.31(0.08) & 0.39(0.07) & 0.85(0.01) \\
101 & 280.71950 & -4.04869 & 1,0,0,0,0 & C$^{18}$O [1] & 78.58(0.08) & 0.34(0.08) & 78.57(0.07) & 0.37(0.07) & 0.40(0.07) & 0.78(0.01) \\
102 & 280.69696 & -4.06383 & 0,1-,1-,0,0 & - & - & - & - & - & - & 0.71(0.01) \\
103 & 280.69424 & -4.06827 & 0,0,0,0,0 & - & - & - & - & - & - & 0.97(0.01) \\
104 & 280.70462 & -4.03363 & 2,0,0,1,0 & C$^{18}$O [2] & 79.56(0.18) & 0.67(0.18) & - & - & 0.70(0.17) & 0.67(0.01) \\
105 & 280.71134 & -4.05637 & 1,0,0,0,0 & C$^{18}$O [1] & 75.90(0.04) & 0.20(0.04) & 75.89(0.05) & 0.21(0.05) & 0.29(0.03) & 0.88(0.01) \\
106 & 280.72497 & -4.03756 & 0,0,1,0,0 & DCO$^+$ [1] & 79.61(0.05) & 0.17(0.05) & 79.63(0.06) & 0.20(0.06) & 0.25(0.04) & 0.65(0.01) \\
107 & 280.69466 & -4.06420 & 3,0,0,0,0 & C$^{18}$O [3] & 80.95(0.11) & 0.58(0.12) & 80.00(0.11) & 0.39(0.20) & 0.61(0.11) & 0.93(0.01) \\
108 & 280.70254 & -4.03908 & 1,1-,0,0,0 & C$^{18}$O [1] & 83.02(0.07) & 0.22(0.07) & 83.02(0.08) & 0.26(0.08) & 0.32(0.05) & 0.91(0.01) \\
109 & 280.71332 & -4.05315 & 1,0,0,0,0 & C$^{18}$O [1] & 80.99(0.07) & 0.67(0.07) & 80.97(0.07) & 0.66(0.07) & 0.70(0.07) & 0.85(0.01) \\
110 & 280.71389 & -4.05138 & 1,0,1,0,0 & DCO$^+$ [1] & 79.44(0.05) & 0.24(0.05) & - & - & 0.33(0.04) & 0.99(0.01) \\
111 & 280.72241 & -4.03605 & 0,0,0,1-,0 & - & - & - & - & - & - & 0.86(0.01) \\
112 & 280.70607 & -4.04224 & 0,0,0,0,0 & - & - & - & - & - & - & 0.65(0.00) \\
113 & 280.71368 & -4.05131 & 1,0,0,0,0 & C$^{18}$O [1] & 78.75(0.08) & 0.36(0.08) & 78.72(0.08) & 0.38(0.08) & 0.42(0.07) & 0.99(0.01) \\
114 & 280.72651 & -4.03976 & 1,0,1-,1,0 & C$^{18}$O [1] & 79.16(0.12) & 0.69(0.12) & 79.39(0.22) & 0.69(0.22) & 0.73(0.11) & 0.78(0.01) \\
115 & 280.71189 & -4.05244 & 0,0,0,1-,0 & - & - & - & - & - & - & 0.90(0.01) \\
116 & 280.71105 & -4.05394 & 1,1,0,0,0 & N$_2$D$^+$ [1] & 79.88(0.08) & 0.28(0.08) & - & - & 0.36(0.06) & 1.03(0.01) \\
117 & 280.71317 & -4.05231 & 0,1,0,0,1- & N$_2$D$^+$ [1] & 78.53(0.09) & 0.42(0.09) & 78.49(0.08) & 0.41(0.08) & 0.47(0.08) & 0.82(0.01) \\
118 & 280.69525 & -4.06920 & 0,1-,0,0,1- & - & - & - & - & - & - & 0.84(0.01) \\
119 & 280.69415 & -4.07099 & 2,1-,1,0,1- & DCO$^+$ [1] & 79.64(0.09) & 0.32(0.09) & 79.60(0.05) & 0.31(0.05) & 0.39(0.07) & 0.92(0.01) \\
120 & 280.71218 & -4.05341 & 2-,0,0,0,0 & C$^{18}$O [2] & 81.48(0.07) & 0.30(0.07) & 81.41(0.08) & 0.30(0.08) & 0.37(0.06) & 1.03(0.01) \\
121 & 280.70971 & -4.03259 & 0,0,0,1,0 & - & - & - & - & - & - & 0.73(0.01) \\
122 & 280.72447 & -4.04312 & 0,0,0,1,0 & - & - & - & - & - & - & 0.74(0.01) \\
123 & 280.69550 & -4.06812 & 1,0,1,1-,0 & DCO$^+$ [1] & 81.38(0.04) & 0.22(0.04) & 81.40(0.05) & 0.22(0.05) & 0.28(0.03) & 0.96(0.01) \\
124 & 280.72401 & -4.04231 & 1,0,1,0,0 & C$^{18}$O [1] & 77.66(0.11) & 0.53(0.11) & 77.69(0.11) & 0.51(0.11) & 0.57(0.10) & 0.85(0.01) \\
125 & 280.72358 & -4.04461 & 2,0,0,0,0 & C$^{18}$O [2] & 77.92(0.20) & 0.64(0.22) & 77.96(0.17) & 0.89(0.17) & 0.67(0.21) & 0.74(0.01) \\
126 & 280.70435 & -4.04142 & 1,0,0,0,0 & C$^{18}$O [1] & - & - & - & - & - & 0.78(0.01) \\
127 & 280.72680 & -4.04176 & 0,2-,0,0,0 & N$_2$D$^+$ [2] & - & - & 81.60(0.07) & 0.22(0.07) & - & 0.82(0.01) \\
128 & 280.69536 & -4.06892 & 1-,0,2,0,0 & - & - & - & - & - & - & 0.92(0.01) \\
129 & 280.70506 & -4.04077 & 2-,0,0,1-,0 & C$^{18}$O [1] & - & - & - & - & - & 0.70(0.01) \\
130 & 280.71070 & -4.05491 & 2,0,0,0,0 & C$^{18}$O [2] & 80.04(0.11) & 0.59(0.11) & 80.12(0.12) & 0.71(0.12) & 0.63(0.10) & 0.87(0.01) \\
131 & 280.71294 & -4.05227 & 2-,1,1-,0,1- & N$_2$D$^+$ [1] & 78.68(0.09) & 0.38(0.09) &  &  & 0.44(0.08) & 0.90(0.01) \\
132 & 280.70446 & -4.04181 & 0,0,0,0,0 & - & - & - & - & - & - & 0.78(0.01) \\
133 & 280.72738 & -4.03713 & 0,0,0,1,1 & - & - & - & - & - & - & 0.92(0.01) \\
134 & 280.72534 & -4.04065 & 1,0,1-,2,1 & - & - & - & - & - & - & 0.87(0.01) \\
135 & 280.70484 & -4.04055 & 3,0,1-,0,1- & C$^{18}$O [2] & 81.55(0.09) & 0.40(0.09) & 81.58(0.05) & 0.22(0.05) & 0.45(0.08) & 0.70(0.01) \\
136 & 280.69426 & -4.06945 & 1-,0,1,0,0 & DCO$^+$ [1] & 76.16(0.09) & 0.25(0.09) & - & - & 0.31(0.07) & 0.85(0.01) \\
137 & 280.72506 & -4.04404 & 0,0,0,0,0 & - & - & - & - & - & - & 0.68(0.01) \\
138 & 280.72371 & -4.04441 & 1,0,0,0,0 & C$^{18}$O [1] & 78.93(0.05) & 0.37(0.05) & 78.85(0.06) & 0.42(0.06) & 0.42(0.04) & 0.81(0.01) \\
139 & 280.72596 & -4.04298 & 0,0,0,0,0 & - & - & - & - & - & - & 0.67(0.01) \\
140 & 280.70971 & -4.03299 & 1,0,0,1,0 & C$^{18}$O [1] & 78.37(0.09) & 0.60(0.09) & 78.33(0.08) & 0.58(0.08) & 0.64(0.09) & 0.73(0.01) \\
141 & 280.72457 & -4.04340 & 1-,0,0,0,0 & - & - & - & - & - & - & 0.74(0.01) \\
142 & 280.70575 & -4.03348 & 0,0,0,0,0 & - & - & - & - & - & - & 0.68(0.01) \\
143 & 280.70989 & -4.05477 & 3-,0,0,0,0 & C$^{18}$O [2] & - & - & 77.99(0.10) & 0.50(0.10) & - & 0.93(0.01) \\
144 & 280.69326 & -4.06758 & 1,0,0,0,0 & C$^{18}$O [1] & - & - & - & - & - & 0.93(0.01) \\
145 & 280.71024 & -4.05505 & 1,0,0,0,1- & C$^{18}$O [1] & 79.97(0.10) & 0.57(0.10) & 79.88(0.15) & 0.85(0.15) & 0.61(0.09) & 0.93(0.01) \\
146 & 280.71216 & -4.05363 & 2,0,0,0,0 & C$^{18}$O [2] & 79.75(0.04) & 0.20(0.04) & 79.76(0.04) & 0.20(0.04) & 0.31(0.03) & 1.17(0.01) \\
147 & 280.71274 & -4.05217 & 2-,0,0,0,0 & C$^{18}$O [1] & - & - & 76.76(0.10) & 0.30(0.10) & - & 0.90(0.01) \\
148 & 280.71109 & -4.05488 & 1,0,0,0,0 & C$^{18}$O [1] & 80.38(0.10) & 0.64(0.10) & 80.37(0.10) & 0.66(0.10) & 0.67(0.10) & 0.91(0.01) \\
149 & 280.70270 & -4.03879 & 0,0,0,0,0 & - & - & - & - & - & - & 0.91(0.01) \\
150 & 280.69566 & -4.06966 & 0,0,0,0,0 & - & - & - & - & - & - & 0.84(0.01) \\
151 & 280.69370 & -4.07165 & 1,0,1,0,0 & C$^{18}$O [1] & - & - & - & - & - & 0.87(0.01) \\
152 & 280.71154 & -4.05366 & 0,0,0,0,0 & - & - & - & - & - & - & 1.15(0.01) \\
153 & 280.70616 & -4.03994 & 0,0,0,0,0 & - & - & - & - & - & - & 0.66(0.01) \\
154 & 280.70448 & -4.03354 & 1-,0,0,2,0 & - & - & - & - & - & - & 0.65(0.01) \\
155 & 280.70412 & -4.04137 & 1,0,0,0,0 & C$^{18}$O [1] & 81.36(0.06) & 0.18(0.06) & 81.37(0.07) & 0.18(0.06) & 0.30(0.04) & 0.77(0.01) \\
156 & 280.72893 & -4.03681 & 0,0,0,0,0 & - & - & - & - & - & - & 0.79(0.01) \\
157 & 280.71251 & -4.05288 & 0,0,0,0,1- & - & - & - & - & - & - & 0.99(0.01) \\
158 & 280.70641 & -4.03617 & 0,0,0,0,0 & - & - & - & - & - & - & 0.71(0.01) \\
159 & 280.69589 & -4.06947 & 2,0,0,0,0 & C$^{18}$O [2] & 79.22(0.07) & 0.22(0.07) & 79.22(0.06) & 0.20(0.06) & 0.27(0.06) & 0.77(0.01) \\
160 & 280.72548 & -4.04319 & 0,0,0,0,0 & - & - & - & - & - & - & 0.77(0.01) \\
161 & 280.70253 & -4.03873 & 0,0,1,0,0 & DCO$^+$ [1] & - & - & - & - & - & 0.91(0.01) \\
162 & 280.72651 & -4.03988 & 1,0,0,0,0 & C$^{18}$O [1] & 79.57(0.14) & 0.94(0.14) & 79.51(0.13) & 0.88(0.13) & 0.97(0.14) & 0.78(0.01) \\
163 & 280.72191 & -4.04186 & 0,0,0,0,0 & - & - & - & - & - & - & 0.62(0.01) \\
164 & 280.69429 & -4.06890 & 1,0,0,0,0 & C$^{18}$O [1] & 77.29(0.06) & 0.36(0.06) & 77.31(0.06) & 0.37(0.06) & 0.40(0.05) & 0.92(0.01) \\
165 & 280.69261 & -4.07167 & 1,0,0,0,0 & C$^{18}$O [1] & - & - & 80.73(0.11) & 0.48(0.11) & - & 0.87(0.01) \\
166 & 280.70367 & -4.03620 & 0,0,0,0,0 & - & - & - & - & - & - & 0.79(0.00) \\
167 & 280.71047 & -4.05587 & 3-,0,1-,0,0 & - & - & - & - & - & - & 0.87(0.01) \\
168 & 280.70971 & -4.05529 & 2,0,0,0,0 & C$^{18}$O [2] & 79.90(0.05) & 0.33(0.05) & 79.91(0.05) & 0.33(0.05) & 0.40(0.04) & 0.92(0.01) \\
169 & 280.71381 & -4.05161 & 1,1-,1,0,0 & C$^{18}$O [1] & 78.67(0.09) & 0.53(0.09) & 78.64(0.09) & 0.52(0.09) & 0.57(0.08) & 0.91(0.01) \\
170 & 280.72795 & -4.01456 & 1,0,0,0,0 & C$^{18}$O [1] & 78.83(0.08) & 0.33(0.08) & 78.86(0.09) & 0.33(0.09) & 0.38(0.07) & 0.75(0.01) \\
171 & 280.70196 & -4.03836 & 0,0,1-,0,0 & - & - & - & - & - & - & 0.85(0.01) \\
172 & 280.69424 & -4.07163 & 2,0,1-,0,0 & C$^{18}$O [2] & 80.90(0.11) & 0.63(0.11) & 80.90(0.09) & 0.61(0.09) & 0.67(0.10) & 1.01(0.01) \\
173 & 280.69385 & -4.06931 & 1,1-,0,0,0 & - & - & - & - & - & - & 0.80(0.01) \\
174 & 280.69514 & -4.06956 & 0,0,0,0,0 & - & - & - & - & - & - & 0.90(0.01) \\
175 & 280.70609 & -4.03341 & 0,0,0,1,0 & - & - & - & - & - & - & 0.72(0.01) \\
176 & 280.69502 & -4.06891 & 1,0,0,1-,1- & C$^{18}$O [1] & 77.20(0.07) & 0.25(0.07) & 77.22(0.06) & 0.24(0.06) & 0.30(0.06) & 0.93(0.01) \\
177 & 280.70974 & -4.05826 & 2,0,0,0,1- & C$^{18}$O [2] & 75.72(0.07) & 0.42(0.07) & 75.73(0.08) & 0.46(0.08) & 0.47(0.06) & 1.03(0.01) \\
178 & 280.71010 & -4.03322 & 1,0,0,0,0 & C$^{18}$O [1] & 77.99(0.08) & 0.46(0.08) & 77.97(0.09) & 0.47(0.09) & 0.51(0.07) & 0.81(0.01) \\
179 & 280.68692 & -4.06890 & 1,0,0,0,0 & C$^{18}$O [1] & 80.10(0.04) & 0.24(0.04) & 80.10(0.04) & 0.25(0.04) & 0.31(0.03) & 0.66(0.01) \\
180 & 280.69486 & -4.06912 & 1-,0,0,0,0 & - & - & - & - & - & - & 0.93(0.01) \\
181 & 280.69341 & -4.06981 & 1,0,1-,1,0 & C$^{18}$O [1] & 76.78(0.14) & 0.52(0.14) & 76.82(0.13) & 0.51(0.13) & 0.55(0.13) & 0.78(0.01) \\
182 & 280.72489 & -4.04101 & 0,0,0,0,0 & - & - & - & - & - & - & 1.06(0.01) \\
183 & 280.69587 & -4.06840 & 0,0,0,0,0 & - & - & - & - & - & - & 0.94(0.01) \\
184 & 280.70570 & -4.05749 & 1,0,0,0,1- & C$^{18}$O [1] & 79.50(0.05) & 0.27(0.05) & 79.50(0.06) & 0.31(0.06) & 0.34(0.04) & 0.77(0.01) \\
185 & 280.71992 & -4.04694 & 1,0,0,0,1 & C$^{18}$O [1] & 79.28(0.06) & 0.61(0.06) & 79.29(0.06) & 0.59(0.06) & 0.64(0.06) & 0.62(0.01) \\
186 & 280.72354 & -4.03809 & 1,0,1,0,0 & DCO$^+$ [1] & 80.52(0.09) & 0.30(0.09) & 80.46(0.07) & 0.27(0.07) & 0.36(0.08) & 0.82(0.01) \\
187 & 280.70765 & -4.05672 & 1,0,1,0,0 & DCO$^+$ [1] & 79.73(0.09) & 0.39(0.09) & 79.74(0.09) & 0.43(0.09) & 0.44(0.08) & 0.95(0.01) \\
188 & 280.72457 & -4.04391 & 1,0,0,0,1 & C$^{18}$O [1] & - & - & - & - & - & 0.72(0.01) \\
189 & 280.70356 & -4.03934 & 1,0,0,0,0 & C$^{18}$O [1] & 77.77(0.04) & 0.22(0.04) & 77.77(0.04) & 0.22(0.04) & 0.32(0.03) & 0.93(0.01) \\
190 & 280.70964 & -4.05658 & 2,0,0,0,0 & C$^{18}$O [2] & 79.23(0.06) & 0.37(0.06) & 79.19(0.07) & 0.41(0.07) & 0.43(0.05) & 0.81(0.01) \\
191 & 280.69534 & -4.06715 & 0,0,0,0,0 & - & - & - & - & - & - & 0.93(0.01) \\
192 & 280.70217 & -4.04013 & 0,0,0,0,1 & - & - & - & - & - & - & 0.83(0.01) \\
193 & 280.69302 & -4.06866 & 0,0,0,1-,0 & - & - & - & - & - & - & 0.83(0.01) \\
194 & 280.70427 & -4.03654 & 0,0,1-,0,0 & - & - & - & - & - & - & 0.78(0.01) \\
195 & 280.71414 & -4.05129 & 1,0,0,0,1- & C$^{18}$O [1] & 78.05(0.17) & 0.73(0.17) & 78.20(0.18) & 0.62(0.20) & 0.76(0.16) & 0.99(0.01) \\
196 & 280.69526 & -4.06872 & 0,0,0,0,0 & - & - & - & - & - & - & 0.92(0.01) \\
197 & 280.68739 & -4.07487 & 1,0,0,0,0 & C$^{18}$O [1] & 80.56(0.02) & 0.19(0.02) & 80.59(0.02) & 0.20(0.02) & 0.27(0.02) & 0.75(0.01) \\
198 & 280.70808 & -4.03304 & 0,0,0,1-,0 & - & - & - & - & - & - & 0.67(0.01) \\
199 & 280.70630 & -4.04229 & 0,0,0,0,0 & - & - & - & - & - & - & 0.65(0.00) \\
200 & 280.68681 & -4.06847 & 1,0,0,0,0 & C$^{18}$O [1] & 80.23(0.13) & 0.54(0.13) & 80.25(0.04) & 0.21(0.04) & 0.58(0.12) & 0.71(0.01) \\
201 & 280.70444 & -4.03619 & 0,0,0,0,0 & - & - & - & - & - & - & 0.78(0.01) \\
202 & 280.69551 & -4.06880 & 0,0,0,0,0 & - & - & - & - & - & - & 0.92(0.01) \\
203 & 280.72506 & -4.04188 & 1,0,1-,0,0 & C$^{18}$O [2] & 81.11(0.07) & 0.33(0.07) & 81.13(0.06) & 0.32(0.06) & 0.40(0.06) & 0.94(0.01) \\
204 & 280.72798 & -4.03917 & 2,0,1,0,0 & C$^{18}$O [1] & 79.39(0.07) & 0.34(0.07) & 79.41(0.05) & 0.30(0.05) & 0.42(0.06) & 0.84(0.01) \\
205 & 280.70392 & -4.03348 & 0,0,0,0,0 & - & - & - & - & - & - & 0.75(0.01) \\
206 & 280.71292 & -4.05247 & 0,0,0,0,0 & - & - & - & - & - & - & 0.90(0.01) \\
207 & 280.70582 & -4.03249 & 0,0,0,0,0 & - & - & - & - & - & - & 0.59(0.01) \\
208 & 280.72609 & -4.04217 & 1,0,0,0,0 & C$^{18}$O [1] & - & - & 81.02(0.04) & 0.16(0.04) & - & 0.64(0.01) \\
209 & 280.69410 & -4.06799 & 1-,0,0,1-,0 & - & - & - & - & - & - & 0.95(0.01) \\
210 & 280.71162 & -4.05270 & 2,0,0,0,1- & C$^{18}$O [2] & 79.12(0.06) & 0.29(0.06) & - & - & 0.37(0.05) & 1.01(0.01) \\
211 & 280.71106 & -4.05376 & 0,0,0,0,0 & - & - & - & - & - & - & 1.03(0.01) \\
212 & 280.69514 & -4.06879 & 1-,0,0,1-,1- & - & - & - & - & - & - & 0.93(0.01) \\
213 & 280.70416 & -4.03777 & 1,0,0,0,0 & C$^{18}$O [1] & 76.75(0.04) & 0.18(0.04) & 76.77(0.04) & 0.18(0.04) & 0.27(0.03) & 0.69(0.01) \\
214 & 280.70783 & -4.05676 & 1,0,0,0,0 & C$^{18}$O [1] & 79.93(0.34) & 1.04(0.34) & 79.87(0.30) & 1.11(0.30) & 1.06(0.33) & 0.95(0.01) \\
215 & 280.72834 & -4.03669 & 0,0,0,0,0 & - & - & - & - & - & - & 0.79(0.01) \\
216 & 280.70330 & -4.03972 & 0,0,1-,0,0 & - & - & - & - & - & - & 0.93(0.01) \\
217 & 280.72653 & -4.04194 & 2-,0,0,0,0 & - & - & - & - & - & - & 0.82(0.01) \\
218 & 280.70409 & -4.03845 & 0,0,1-,0,0 & - & - & - & - & - & - & 0.84(0.01) \\
219 & 280.71336 & -4.05233 & 1,1,0,0,0 & N$_2$D$^+$ [1] & 78.48(0.09) & 0.28(0.09) & - & - & 0.35(0.07) & 0.82(0.01) \\
220 & 280.69537 & -4.07001 & 0,0,1-,0,0 & - & - & - & - & - & - & 0.72(0.01) \\
221 & 280.69281 & -4.06859 & 0,0,0,0,0 & - & - & - & - & - & - & 0.91(0.01) \\
222 & 280.70886 & -4.05570 & 0,0,0,0,0 & - & - & - & - & - & - & 0.97(0.01) \\
223 & 280.69441 & -4.07109 & 1,0,1-,0,1- & C$^{18}$O [1] & 81.10(0.15) & 0.59(0.15) & 81.18(0.14) & 0.62(0.14) & 0.63(0.14) & 0.92(0.01) \\
224 & 280.70558 & -4.04129 & 0,0,0,0,0 & - & - & - & - & - & - & 0.68(0.00) \\
225 & 280.72467 & -4.04354 & 0,0,0,0,0 & - & - & - & - & - & - & 0.74(0.01) \\
226 & 280.71960 & -4.04899 & 0,1,1-,0,0 & N$_2$D$^+$ [1] & 78.85(0.09) & 0.22(0.09) & - & - & 0.30(0.07) & 0.78(0.01) \\
227 & 280.70582 & -4.03236 & 0,0,0,0,0 & - & - & - & - & - & - & 0.59(0.01) \\
228 & 280.72191 & -4.04623 & 0,0,0,0,0 & - & - & - & - & - & - & 0.67(0.01) \\
229 & 280.69454 & -4.06912 & 1,0,0,0,0 & C$^{18}$O [1] & 77.38(0.05) & 0.26(0.05) & 77.39(0.05) & 0.26(0.05) & 0.31(0.04) & 0.92(0.01) \\
230 & 280.72534 & -4.04122 & 1,0,0,0,0 & C$^{18}$O [1] & - & - & - & - & - & 0.86(0.01) \\
231 & 280.69469 & -4.06731 & 0,0,0,0,0 & - & - & - & - & - & - & 0.94(0.01) \\
232 & 280.68728 & -4.06901 & 1,0,1-,0,0 & C$^{18}$O [1] & 80.08(0.05) & 0.24(0.05) & 80.08(0.04) & 0.23(0.04) & 0.31(0.04) & 0.66(0.01) \\
233 & 280.71008 & -4.05663 & 2,0,1,0,0 & DCO$^+$ [1] & 78.82(0.08) & 0.31(0.08) & 78.86(0.07) & 0.29(0.07) & 0.38(0.07) & 0.85(0.01) \\
234 & 280.72429 & -4.04191 & 1,1-,1-,0,0 & C$^{18}$O [1] & 80.99(0.05) & 0.23(0.05) & 81.00(0.05) & 0.23(0.05) & 0.30(0.04) & 0.92(0.01) \\
235 & 280.68646 & -4.06830 & 1-,0,0,0,0 & - & - & - & - & - & - & 0.71(0.01) \\
236 & 280.70438 & -4.03340 & 0,0,0,0,0 & - & - & - & - & - & - & 0.65(0.01) \\
237 & 280.69473 & -4.06922 & 1-,0,0,0,0 & - & - & - & - & - & - & 0.90(0.01) \\
238 & 280.72186 & -4.04202 & 1,0,0,0,0 & - & - & - & - & - & - & 0.64(0.01) \\
239 & 280.72503 & -4.04309 & 2-,0,0,0,0 & C$^{18}$O [1] & 75.60(0.05) & 0.18(0.05) & - & - & 0.25(0.04) & 0.71(0.01) \\
240 & 280.70735 & -4.03547 & 1,0,0,0,0 & C$^{18}$O [1] & - & - & 78.60(0.12) & 0.44(0.12) & - & 0.63(0.01) \\
241 & 280.70995 & -4.05666 & 3,0,1-,0,0 & C$^{18}$O [3] & 79.49(0.07) & 0.39(0.07) & 79.44(0.07) & 0.34(0.08) & 0.45(0.06) & 0.85(0.01) \\
242 & 280.70148 & -4.03781 & 0,0,0,0,0 & - & - & - & - & - & - & 0.73(0.01) \\
243 & 280.69331 & -4.07116 & 1,0,0,1,0 & C$^{18}$O [1] & 81.12(0.04) & 0.29(0.04) & 81.14(0.04) & 0.31(0.04) & 0.35(0.03) & 0.83(0.01) \\
244 & 280.70299 & -4.03424 & 0,0,0,0,0 & - & - & - & - & - & - & 0.59(0.01) \\
245 & 280.72696 & -4.04136 & 2,0,0,1-,0 & C$^{18}$O [1] & - & - & 81.15(0.04) & 0.18(0.04) & - & 0.85(0.01) \\
246 & 280.72936 & -4.01581 & 0,0,0,0,0 & - & - & - & - & - & - & 0.54(0.01) \\
247 & 280.69481 & -4.07031 & 0,1,0,0,0 & N$_2$D$^+$ [1] & 79.99(0.09) & 0.23(0.09) & - & - & 0.30(0.07) & 0.87(0.01) \\
248 & 280.72717 & -4.01756 & 1,0,0,0,0 & C$^{18}$O [1] & - & - & 78.34(0.08) & 0.30(0.08) & - & 0.93(0.03) \\
249 & 280.70622 & -4.04013 & 0,0,1,0,0 & DCO$^+$ [1] & - & - & 80.01(0.09) & 0.45(0.09) & - & 0.66(0.01) \\
250 & 280.69310 & -4.06845 & 0,0,0,0,0 & - & - & - & - & - & - & 0.86(0.01) \\
251 & 280.72896 & -4.03881 & 0,0,0,0,0 & - & - & - & - & - & - & 0.76(0.01) \\
252 & 280.69526 & -4.06837 & 1,0,0,0,1- & C$^{18}$O [1] & - & - & 77.48(0.07) & 0.28(0.07) & - & 0.96(0.01) \\
253 & 280.69369 & -4.06826 & 0,0,0,1-,0 & - & - & - & - & - & - & 0.90(0.01) \\
254 & 280.69551 & -4.06830 & 0,0,1-,0,0 & - & - & - & - & - & - & 0.96(0.01) \\
255 & 280.70818 & -4.03424 & 0,0,0,0,0 & - & - & - & - & - & - & 0.63(0.01) \\
256 & 280.70622 & -4.03879 & 0,0,1,1-,0 & DCO$^+$ [1] & 79.74(0.08) & 0.32(0.08) & 79.74(0.07) & 0.33(0.07) & 0.39(0.07) & 0.71(0.01) \\
257 & 280.68661 & -4.07547 & 1,0,0,0,0 & C$^{18}$O [1] & 80.88(0.03) & 0.19(0.03) & 80.88(0.03) & 0.19(0.03) & 0.25(0.03) & 0.85(0.02) \\
258 & 280.70433 & -4.04205 & 0,1,0,0,0 & N$_2$D$^+$ [1] & 79.86(0.12) & 0.33(0.12) & 79.90(0.13) & 0.38(0.13) & 0.40(0.10) & 0.80(0.01) \\
259 & 280.72592 & -4.01434 & 2,0,0,0,0 & C$^{18}$O [1] & 78.26(0.04) & 0.22(0.04) & 78.26(0.04) & 0.23(0.04) & 0.31(0.03) & 0.93(0.02) \\
260 & 280.70323 & -4.04172 & 0,0,1,0,0 & DCO$^+$ [1] & 81.25(0.12) & 0.39(0.12) & 81.18(0.14) & 0.43(0.14) & 0.45(0.11) & 0.75(0.01) \\
261 & 280.71356 & -4.05261 & 2,0,0,1-,1- & C$^{18}$O [2] & 78.69(0.10) & 0.54(0.10) & 78.69(0.10) & 0.55(0.10) & 0.58(0.09) & 0.80(0.01) \\
262 & 280.69569 & -4.06987 & 0,0,1,0,0 & DCO$^+$ [1] & - & - & 79.53(0.07) & 0.22(0.07) & - & 0.72(0.01) \\
263 & 280.69292 & -4.07190 & 0,0,0,0,0 & - & - & - & - & - & - & 0.87(0.01) \\
264 & 280.70258 & -4.03981 & 0,0,0,0,0 & - & - & - & - & - & - & 0.85(0.01) \\
265 & 280.70584 & -4.05745 & 1,0,0,0,0 & C$^{18}$O [1] & - & - & 79.40(0.04) & 0.14(0.03) & - & 0.89(0.01) \\
266 & 280.72685 & -4.04159 & 0,0,0,1-,0 & - & - & - & - & - & - & 0.82(0.01) \\
267 & 280.72691 & -4.04187 & 1,0,0,0,0 & C$^{18}$O [1] & 77.59(0.06) & 0.22(0.06) & 77.59(0.07) & 0.24(0.07) & 0.31(0.04) & 0.82(0.01) \\
268 & 280.73050 & -4.01813 & 0,0,0,0,0 & - & - & - & - & - & - & 0.44(0.01) \\
269 & 280.72181 & -4.04616 & 0,0,0,0,0 & - & - & - & - & - & - & 0.67(0.01) \\
270 & 280.69351 & -4.07173 & 2,0,0,0,0 & C$^{18}$O [2] & 78.21(0.05) & 0.18(0.05) & - & - & 0.26(0.04) & 0.90(0.01) \\
271 & 280.72520 & -4.04212 & 1,0,0,0,1- & C$^{18}$O [1] & 80.40(0.19) & 0.88(0.19) & 80.37(0.18) & 0.89(0.18) & 0.91(0.18) & 0.88(0.01) \\
272 & 280.69303 & -4.07161 & 2,0,1,0,1- & DCO$^+$ [1] & - & - & 81.39(0.05) & 0.19(0.04) & - & 0.90(0.01) \\
273 & 280.69441 & -4.06906 & 1,0,0,0,0 & C$^{18}$O [1] & 77.20(0.06) & 0.29(0.06) & 77.22(0.06) & 0.29(0.06) & 0.34(0.05) & 0.92(0.01) \\
274 & 280.68663 & -4.06819 & 1-,0,0,0,0 & - & - & - & - & - & - & 0.71(0.01) \\
275 & 280.69437 & -4.06795 & 1-,0,0,1-,1- & - & - & - & - & - & - & 1.00(0.01) \\
276 & 280.69646 & -4.06790 & 1,0,0,0,1- & C$^{18}$O [1] & - & - & 77.56(0.09) & 0.32(0.09) & - & 0.93(0.01) \\
277 & 280.70909 & -4.05529 & 2,0,0,0,1 & C$^{18}$O [2] & 79.60(0.07) & 0.38(0.07) & 79.55(0.07) & 0.37(0.07) & 0.45(0.06) & 0.97(0.01) \\
278 & 280.69243 & -4.07195 & 2,0,0,0,0 & C$^{18}$O [2] & 80.60(0.11) & 0.48(0.11) & 80.81(0.12) & 0.42(0.14) & 0.52(0.10) & 0.89(0.01) \\
279 & 280.69359 & -4.06930 & 0,0,0,0,0 & - & - & - & - & - & - & 0.80(0.01) \\
280 & 280.70541 & -4.05736 & 1,0,0,0,0 & C$^{18}$O [1] & 80.28(0.23) & 0.90(0.23) & - & - & 0.92(0.22) & 0.77(0.01) \\
\enddata
\tablecomments{Column (4) lists the line detection for C$^{18}$O,N$_2$D$^+$,DCO$^+$,DCN,CH$_3$OH.}
\end{deluxetable*}

\bibliography{ref}
\bibliographystyle{aasjournal}

\end{document}